\documentclass[twocolappendix]{emulateapj}
\usepackage{subfigure}
\usepackage{amsmath}
\usepackage{color}

\newcommand{\bmath}[1]{\mbox{\boldmath{$#1$}}}

\newcommand{\del}{{\bf \nabla}}

\newcommand{\cs}{c_{\rm s}}

\begin{document}

\title{The Mass and Size Distribution of Planetesimals Formed by the \\ Streaming Instability.
I. The role of Self-Gravity}

\author{Jacob B. Simon\altaffilmark{1,2,3}, Philip J. Armitage\altaffilmark{3,4}, Rixin, Li\altaffilmark{5}, Andrew N. Youdin\altaffilmark{5}}

\email{jbsimon.astro@gmail.com}

\begin{abstract}
We study the formation of planetesimals in protoplanetary disks from the gravitational collapse of solid 
over-densities generated via the streaming instability.
To carry out these studies, we implement and test a particle-mesh self-gravity module for the {\sc Athena} 
code that enables the simulation of aerodynamically coupled systems of gas and collisionless self-gravitating 
solid particles.  Upon employment of our algorithm to planetesimal formation simulations, we find that (when a direct comparison is possible) the {\sc Athena} simulations yield predicted planetesimal properties that agree well with those found in prior work using different numerical techniques. In particular, 
the gravitational collapse of streaming-initiated clumps leads to an initial planetesimal mass function 
that is well-represented by a power-law, ${\rm d}N / {\rm d}M_p \propto M_p^{-p}$, with $p \simeq 1.6 \pm 0.1$, which
equates to a differential size distribution ${\rm d}N / {\rm d}R_p \propto R_p^{-q}$, with $q \simeq  2.8 \pm 0.1$.
We find no significant trends with resolution from a convergence study of up to $512^3$ grid zones and 
$N_{\rm par} \approx 1.5 \times 10^8$ particles. Likewise, the power-law slope appears indifferent to changes in 
the relative strength of self-gravity and tidal shear, and to the time when (for reasons of numerical 
economy) self-gravity is turned on, though the strength of these claims is limited by small number 
statistics. For a typically assumed radial distribution of minimum mass solar nebula solids (assumed here to have dimensionless 
stopping time $\tau = 0.3$), our results support the hypothesis that bodies on the scale of large 
asteroids or Kuiper Belt Objects could have formed as the high-mass tail of a primordial 
planetesimal population.
\end{abstract} 

\keywords{planets and satellites: formation --- hydrodynamics --- instabilities --- turbulence --- 
protoplanetary disks} 

\altaffiltext{1}{Department of Space Studies, Southwest Research Institute, Boulder, CO 80302}
\altaffiltext{2}{Sagan Fellow}
\altaffiltext{3}{JILA, University of Colorado and NIST, 440 UCB, Boulder, CO 80309-0440}
\altaffiltext{4}{Department of Astrophysical and Planetary Sciences, University of Colorado, Boulder, CO 80309-0391}
\altaffiltext{5}{Department of Astronomy and Steward Observatory, University of Arizona, 933 North Cherry Avenue, Tucson, AZ 85721}


\section{Introduction} 
The discovery of thousands of exoplanet systems from {\em Kepler} and other missions has 
confirmed the ubiquity and diversity of planetary systems in our galaxy. The dominant physical processes that 
lead to the observed configurations of exoplanet systems, however, remain unclear. A central question, for 
both high and low-mass planets, is whether what we see reflects the in situ growth of planets from 
a population of planetesimals, or is instead determined largely by migration at a later stage. Answering this 
question robustly requires determining, from protoplanetary disk initial conditions, where and when 
planetesimals form.

In the ``bottom-up" model for planet formation, planets are built through several key stages, beginning with the coagulation of small particles into larger particles and particle aggregates, the formation of planetesimals from these solids, and the growth of planetesimals from accretion into larger bodies that become terrestrial planets and the cores of gas giants.  In the first step, the collisions of particles
through Brownian and turbulent motions lead to their growth into larger bodies.  This process is efficient at producing solids of sizes $\sim$mm \cite[e.g.,][]{brauer08,birnstiel10,zsom10}. Recent work has shown that it may be possible
for solids to grow to $\sim$cm-m sizes as they drift through radially varying turbulence \citep{drazkowska13}, though
interior to the snow line, such growth is difficult and solids likely remain stunted at mm sizes \citep{drazkowska14}.

The propensity of these solids to grow in size only goes so far, however, and forming larger bodies of km size scales or larger (i.e., planetesimal scales) faces two theoretical difficulties. 
First, if particles grow to sufficiently large sizes ($\sim$ meter sizes at 1 AU), they experience
a significant head-wind from the sub-Keplerian rotating gas, which causes them to lose angular momentum and rapidly spiral into the star \citep{weidenschilling77b}.  Several mechanisms
have been invoked to counteract this drift.  For example, pressure bumps due to zonal flows \citep{johansen09a,simon14}, ice lines 
(\citealt{kretke07}; but see \citealt{yang10} and \citealt{bitsch14}), and abrupt transitions in ionization fraction \citep{dzyurkevich10,drazkowska13} may stop or slow this radial drift and may even be required given observations showing the presence of these particles at large disk radii \cite[e.g.,][]{andrews12}.\footnote{Though, as shown by \cite{birnstiel12}, the presence of small grains at large radii may be the result of very long timescales over which these grains are swept up by larger solids.}
However, even if these particle traps are efficient at slowing radial drift, particles still cannot easily grow beyond mm-cm sizes. A combination of laboratory experiments (primarily on silicates) and modeling shows that particles within this size range do not easily stick together, and instead fragment or bounce \citep{brauer08,blum08,birnstiel10,zsom10}. Whether this is also the case for icy particles is less clear \cite[e.g.,][]{blum08,okuzumi12,kataoka13,wada13,krijt15,musiolik16}.

A promising route toward surpassing these difficulties involves instabilities of aerodynamically 
coupled systems of gas and solid particles.  As the gas removes angular momentum from the particles via a headwind, the particles experience inward radial drift. However, the back-reaction of the particle momentum on the gas causes particle over-densities to experience less gas drag as these over-densities act to boost the gas and reduce their own headwind. Therefore, regions of higher particle density will experience slower radial drift, leading to a pile up of solids at various radii. This mechanism is the essence of what is known as the streaming instability \citep{youdin05,youdin07a,johansen07b,bai10b,bai10c}.

The streaming instability has been studied analytically in the linear regime \citep{youdin05,youdin07a} and in the non-linear regime via numerical simulations \cite[e.g.,][]{johansen07a,johansen07b,bai10c}. The initial 
growth of the instability does not involve self-gravity, and in this limit, two and three-dimensional calculations have 
been used to quantify the strength of clumping as a function of the physical \citep{johansen09c,bai10b,carrera15}, 
and numerical parameters, e.g., resolution \citep{bai10a}. In particular, \cite{johansen09c} and \cite{bai10b} carried out numerical calculations of the streaming instability
without self-gravity and found that the instability generally leads to efficient particle clumping when the dimensionless stopping time (i.e., the stopping time multiplied by the orbital frequency) is
$\tau \gtrsim 10^{-2}$ and the height-integrated solid-to-gas ratio is super-solar.  Furthermore, the instability occurs for a wide range in background gas pressure gradients, but for a given metallicity, it favors
smaller gradients \citep{bai10b}. 

Three-dimensional calculations that include the mutual gravitational forces between particles have been used 
to study how planetesimals form \citep{johansen07a,johansen09c,johansen11a,johansen12,johansen15}.  While many characteristics of planetesimal formation have yet to be fully explored, these studies have generally shown that planetesimals with masses consistent with dwarf planets and planetesimals in the main asteroid belt and Kuiper belt can form after the streaming instability generates sufficiently strong clumps that self-gravity can take over \cite[e.g.,][]{johansen07a,johansen11a,johansen12,johansen15}.  

These studies have established the streaming instability as the leading candidate mechanism for the 
efficient formation of planetesimals. Many critical questions, however, remain open. These include whether 
the streaming instability --- in isolation, or in concert with large-scale structures such as zonal flows \citep{johansen09a} --- can 
form planetesimals in the inner disk, where the dimensionless stopping time of mm-sized particles 
$\tau \ll 1$. Furthermore, the dependence of the initial planetesimal
mass and size distributions on physical properties, such as $\tau$, the metallicity, and gas-phase turbulence
is an open issue. Indeed, characterizing these distributions is of substantial importance to explaining properties of both the main asteroid belt and the Kuiper belt planetesimal populations. 
To-date, simulations of the streaming instability in the presence of particle self-gravity \citep{johansen15} have revealed a shallower planetesimal size distribution than what is inferred from main belt and Kuiper belt observations \citep{jedicke02,morbidelli09,fraser10,fraser14}. However,
an extensive systematic survey of the dependence of this distribution on physical parameters has yet to be carried out. 

Purely numerical issues are also of interest. The relative 
strengths and weakness of different numerical schemes are well-characterized in a variety of situations, 
including low and high-Mach number turbulence, but much less is known in the more complex situation 
where we have multiple phases, aerodynamic coupling and self-gravity. All previous simulations of 
planetesimal formation that have included self-gravity have used high-order finite difference methods 
(implemented in the {\sc Pencil} code). Here, we instead combine particle self-gravity 
with a higher order Godunov hydrodynamic scheme \cite[implemented in the {\sc Athena} code;][]{stone08} to 
study planetesimal formation and again address the dependence of planetesimal mass and size distributions
on numerical effects.

In this paper, the first of a series, we describe our implementation of particle self-gravity within the {\sc Athena} 
code, and present results from a baseline set of 3D simulations. Our primary focus is on independently 
verifying (and in some cases extending) prior numerical simulations of streaming-initiated 
planetesimal formation, and hence we start with parameter choices that match those in the existing 
literature. We specifically address the convergence of the initial planetesimal mass function with numerical resolution, and 
the effect of varying the strength of self-gravity as compared to tidal shear. We also study whether 
the results depend on {\em when} in the simulation particle self-gravity is turned on, given that it 
is common (and computationally expedient) to do so at a relatively late time when the non-self-gravitating 
streaming instability is already strongly non-linear. There has been some study of the impact of this 
approximation \citep{johansen11a}, but a detailed investigation of the effects of pre-gravity conditions on the outcome of planetesimal formation has not yet been performed.

The outline of the paper is as follows. Our methodology is described in detail in Section~\ref{method}, which includes a description of the numerical algorithm and the implementation of particle self-gravity, two test problems 
to check this implementation, and a description of our streaming instability calculations.  In Section~\ref{results}, we present our results from each set of calculations, and we discuss these
results in Section~\ref{discussion}.  We wrap up with a summary and conclusions in Section~\ref{conclusions}.


\begin{deluxetable*}{l|cccccccc}
\tabletypesize{\small}
\tablewidth{0pc}
\tablecaption{Streaming Instability Simulations\label{tbl:sims}}
\tablehead{
\colhead{Run}&
\colhead{Domain Size}&
\colhead{Resolution}&
\colhead{$N_{\rm par}$}&
\colhead{$\tau$}&
\colhead{$Z$}&
\colhead{$\tilde{G}$}&
\colhead{$t_{\rm sg}$}&
\colhead{Comments} \\
\colhead{ }&
\colhead{$(L_x\times L_y\times L_z)H$}&
\colhead{$N_x\times N_y\times N_z$}&
\colhead{ }&
\colhead{ }&
\colhead{ }&
\colhead{ }&
\colhead{($\Omega^{-1}$)}&
\colhead{ } } 
\startdata
SI64-G0.05 & $0.2\times0.2\times0.2$ & $64\times64\times64$ & 300,000 & 0.3 & 0.02 & 0.05 & 400 & --\\
SI128-G0.05 & $0.2\times0.2\times0.2$ & $128\times128\times128$ & 2,400,000 & 0.3 & 0.02 & 0.05 & 170 & Fiducial Run \\
SI256-G0.05 & $0.2\times0.2\times0.2$ & $256\times256\times256$ & 19,200,000 & 0.3 & 0.02 & 0.05 & 150 & -- \\
SI512-G0.05 & $0.2\times0.2\times0.2$ & $512\times512\times512$ & 153,600,000 & 0.3 & 0.02 & 0.05 & 110 & -- \\
\vspace{0.003in} \\
\hline \\
SI128-G0.02 & $0.2\times0.2\times0.2$ & $128\times128\times128$ & 2,400,000 & 0.3 & 0.02 & 0.02 & 170 & -- \\
SI128-G0.02\_tm20 & $0.2\times0.2\times0.2$ & $128\times128\times128$ & 2,400,000 & 0.3 & 0.02 & 0.02 & 150 & Restarted $20\Omega^{-1}$ earlier \\
SI128-G0.02\_tm10 & $0.2\times0.2\times0.2$ & $128\times128\times128$ & 2,400,000 & 0.3 & 0.02 & 0.02 & 160 & Restarted $10\Omega^{-1}$ earlier  \\
SI128-G0.02\_tp10 & $0.2\times0.2\times0.2$ & $128\times128\times128$ & 2,400,000 & 0.3 & 0.02 & 0.02 & 180 & Restarted $10\Omega^{-1}$ later  \\
SI128-G0.02\_tp20 & $0.2\times0.2\times0.2$ & $128\times128\times128$ & 2,400,000 & 0.3 & 0.02 & 0.02 & 190 & Restarted $20\Omega^{-1}$ later  \\
SI128-G0.05 & $0.2\times0.2\times0.2$ & $128\times128\times128$ & 2,400,000 & 0.3 & 0.02 & 0.05 & 170 & Fiducial Run \\
SI128-G0.05\_tm20 & $0.2\times0.2\times0.2$ & $128\times128\times128$ & 2,400,000 & 0.3 & 0.02 & 0.05 & 150 & Restarted $20\Omega^{-1}$ earlier \\
SI128-G0.05\_tm10 & $0.2\times0.2\times0.2$ & $128\times128\times128$ & 2,400,000 & 0.3 & 0.02 & 0.05 & 160 & Restarted $10\Omega^{-1}$ earlier  \\
SI128-G0.05\_tp10 & $0.2\times0.2\times0.2$ & $128\times128\times128$ & 2,400,000 & 0.3 & 0.02 & 0.05 & 180 & Restarted $10\Omega^{-1}$ later  \\
SI128-G0.05\_tp20 & $0.2\times0.2\times0.2$ & $128\times128\times128$ & 2,400,000 & 0.3 & 0.02 & 0.05 & 190 & Restarted $20\Omega^{-1}$ later  \\
SI128-G0.1 & $0.2\times0.2\times0.2$ & $128\times128\times128$ & 2,400,000 & 0.3 & 0.02 & 0.1 & 170 & -- \\
SI128-G0.1\_tm20 & $0.2\times0.2\times0.2$ & $128\times128\times128$ & 2,400,000 & 0.3 & 0.02 & 0.02 & 150 & Restarted $20\Omega^{-1}$ earlier \\
SI128-G0.1\_tm10 & $0.2\times0.2\times0.2$ & $128\times128\times128$ & 2,400,000 & 0.3 & 0.02 & 0.1 & 160 & Restarted $10\Omega^{-1}$ earlier  \\
SI128-G0.1\_tp10 & $0.2\times0.2\times0.2$ & $128\times128\times128$ & 2,400,000 & 0.3 & 0.02 & 0.1 & 180 & Restarted $10\Omega^{-1}$ later  \\
SI128-G0.1\_tp20 & $0.2\times0.2\times0.2$ & $128\times128\times128$ & 2,400,000 & 0.3 & 0.02 & 0.1 & 190 & Restarted $20\Omega^{-1}$ later  \\
\vspace{0.003in} \\
\hline \\
SI128-G0.05-no\_clump & $0.2\times0.2\times0.2$ & $128\times128\times128$ & 2,400,000 & 0.3 & 0.02 & 0.05 & 0 & --  \\
SI128-G0.05-low\_clump & $0.2\times0.2\times0.2$ & $128\times128\times128$ & 2,400,000 & 0.3 & 0.02 & 0.05 &40 & -- \\
SI128-G0.05-med\_clump & $0.2\times0.2\times0.2$ & $128\times128\times128$ & 2,400,000 & 0.3 & 0.02 & 0.05 & 170 & {\footnotesize Same as Fiducial Run} \\
SI128-G0.05-high\_clump & $0.2\times0.2\times0.2$ & $128\times128\times128$ & 2,400,000 & 0.3 & 0.02 & 0.05 & 240 &  -- \\
\enddata
\end{deluxetable*}


\section{Method}
\label{method}

In this section, we explain our methodology.  We first describe the algorithmic details of 
{\sc Athena} in the shearing box approximation with the inclusion of aerodynamic coupling
between the gas and the particles and the mutual gravitational attraction between particles
solved via a Fast Fourier Transform (FFT) method.  We
then present tests of our particle self-gravity module, followed by a description of 
our set up, parameters, and diagnostics for the local, streaming instability simulations
that we carry out in this paper.

\subsection{Numerical Algorithm}

Our simulations use {\sc Athena}, a second-order accurate Godunov
flux-conservative code for solving the equations of hydrodynamics and magnetohydrodynamics.  
The simulations we have carried out here neglect magnetic fields, and we use the hydrodynamics module of {\sc Athena}.
We use the {\sc Athena} configuration that includes the dimensionally unsplit corner transport upwind method
of \cite{colella90} coupled with the third-order in space piecewise
parabolic method of \cite{colella84}. We use the HLLC Riemann solver to calculate the
numerical fluxes \citep{toro99}.  A detailed description
of the base {\sc Athena} algorithm and the results of various test problems
are given in \cite{gardiner05a}, \cite{gardiner08}, and \cite{stone08}.

The simulations employ a local shearing box approximation.
The shearing box models a co-rotating disk patch whose size is small compared to the
radial distance from the central object, $R_0$.  This allows the
construction of a local Cartesian frame $(x,y,z)$ that is defined in terms of the disk's
cylindrical co-ordinates $(R,\phi,z^\prime)$ via  $x=(R-R_0)$, $y=R_0 \phi$, and $z = z^\prime$.
The local patch  co-rotates with an angular velocity $\Omega$ corresponding to
the orbital frequency at $R_0$, the center of the box; see \cite{hawley95a}.  

There are two sets of equations to solve.  Comprising the first set are the continuity and momentum equation for the gas dynamics:

\begin{equation}
\label{continuity_eqn}
\frac{\partial \rho}{\partial t} + \del \cdot (\rho {\bmath u}) = 0,
\end{equation}

\begin{eqnarray}
\label{momentum_eqn}
\frac{\partial \rho {\bmath u}}{\partial t} + \del \cdot \left(\rho {\bmath u}{\bmath u} + P {\bmath I} \right) 
& = & 2 q \rho \Omega^2 {\bmath x} - \rho \Omega^2 {\bmath z} \nonumber \\
& & -2{\bmath \Omega} \times \rho {\bmath u} + \rho_p \frac{{\bmath v}-{\bmath u}}{t_{\rm stop}}
\end{eqnarray}

\noindent 
where $\rho$ is the mass density, $\rho {\bmath u}$ is the momentum
density, $P$ is the gas pressure, ${\bmath I}$ is the identity matrix, $t_{\rm stop}$ is the (dimensional) stopping time of the particles,
and $q$ is the shear parameter, defined as $q = -d$ln$\Omega/d$ln$R$.  
We use $q = 3/2$, appropriate for a Keplerian disk.  For simplicity and numerical convenience, we
assume an isothermal equation of state $P = \rho \cs^2$, where $\cs$
is the isothermal sound speed.  From left to right, the source terms
in equation~(\ref{momentum_eqn}) correspond to radial tidal forces
(gravity and centrifugal), vertical gravity, the Coriolis force, and the feedback from the particle momentum onto the gas. 
 The feedback term consists of the local particle mass density $\rho_{\rm p}$, the difference between the particle velocity ${\bmath v}$ and gas velocity, and the stopping time $t_{\rm stop}$ of particles due to gas drag.  As we describe below, this feedback term is calculated at the location of every particle and then distributed to the gas grid points. 

The second set of equations describes the particle evolution. The equation of motion for particle $i$ is given by

\begin{eqnarray}
\label{particle_motion}
\frac{d {\bmath v^\prime_i}}{dt} = 2\left( v^\prime_{iy} - \eta v_{\rm K}\right)& & \Omega \hat{\bmath x} - \left(2 - q\right) v^\prime_{ix} \Omega \hat{\bmath y} \nonumber \\ 
& & - \Omega^2 z \hat{\bmath z} - \frac{{\bmath v^\prime_i} - {\bmath u^\prime}}{t_{\rm stop}} + {\bmath F_{\rm g}}
\end{eqnarray}

\noindent
where the prime denotes a frame in which the background shear velocity has been subtracted.  This is part of the orbital advection scheme \citep{stone10,masset00,johnson08}, which has been implemented for the gas dynamics as well.   The $\eta v_{\rm K}$ term accounts for the inward radial drift of particles resulting from a gas headwind, where $\eta$ is the fraction of the Keplerian velocity by which the orbital velocity of particles is reduced (see Section~\ref{setup}). In real disks, this headwind results from a radial pressure gradient that causes the gas to orbit at sub-Keplerian speeds while the particles continue to orbit at Keplerian velocities. However, such a radial pressure gradient is inconsistent with the radial (shearing) periodic boundary conditions in the shearing box model.   Following \cite{bai10a}, we circumvent this issue by imposing an inward force on the particles, resulting in the $\eta v_{\rm K}$ term as described above. As a result, both the particles and gas (azimuthal) velocities are shifted to slightly higher values (by $\eta v_{\rm K}$) than what would be present in a real disk, but the essential physics of differential motion between the gas and particles is accurately captured.

Equation~(\ref{particle_motion}) is solved using the algorithms described in \cite{bai10a}; in particular, we use the semi-implicit integration method combined with a triangular shaped cloud 
(TSC) scheme to map the particle momentum feedback to the grid cell centers and inversely
to interpolate the gas velocity to the particle locations (${\bmath u^\prime}$). 

The force due to the particle self-gravity is denoted by ${\bmath F_{\rm g}}$, which is found by first solving Poisson's equation for particle self-gravity,

\begin{equation}
\label{poisson}
\del^2 \Phi_{\rm p} = 4 \pi G \rho_{\rm p}
\end{equation}

\noindent
where $\Phi_{\rm p}$ is the gravitational potential of the particle self-gravity.  The force is then calculated via,

\begin{equation}
\label{grav_force}
{\bmath F_{\rm g}} = -{\bmath \del} \Phi_{\rm p}.
\end{equation}

In solving Equation~(\ref{poisson}), we follow the methodology outlined in Section 2.3 and the Appendix of \cite{koyama09}, which consists of a discrete 3D FFT adapted to handle the shearing-periodic boundaries in $x$ and the vacuum boundaries in $z$.  Solving the Poisson equation with FFT requires periodicity in all three dimensions. In the azimuthal dimension, this is trivially satisfied.  For the radial boundaries, the particle mass density (which is distributed to the gas grid points using the same interpolation method as used to calculate the momentum exchange terms; TSC) is mapped to the nearest time in which the shearing-periodic boundaries are purely periodic.  From \cite{hawley95a}, the times at which the radial boundaries are purely periodic are $t_n = n L_y/(q\Omega L_x)$ with $n = 0, 1, 2, 3 ... \ $ .  For each value of $x$, the particle density is reconstructed along $y$ via a conservative remap; the density is calculated by differencing numerical ``fluxes" along $y$.  These fluxes are calculated via third order reconstruction coupled with the extremum preserving algorithm of \cite{sekora09}.   Note that this reconstruction is the same method employed to remap the fluid quantities in the radial ghost zones as part of the shearing-periodic boundary conditions \citep{stone10}.

In the vertical direction, the solution to Equation~(\ref{poisson}) is found using the Green's function of the Poisson equation for a horizontal sheet of sinusoidal source mass, as described in detail in the Appendix
of \cite{koyama09}.  The authors of that work developed this method to calculate the potential of self-gravitating gas, but we have trivially extended this same algorithm to use the particle mass density instead of the gas density.

With these methods in hand, the potential is calculated as follows. The particle mass density is reconstructed along $y$ via the same third-order conservative remap as is used for the boundary conditions, multiplied by appropriate functions of $z$, as described above \cite[see Equation (A11) of][]{koyama09}, and then Fourier transformed via a 3D FFT.  The density in Fourier space is then multiplied by the appropriate coefficients and transformed back to real space via another 3D FFT \cite[see Equation (A8) of][]{koyama09}.  In calculating the 3D FFT's, a domain of size  $L_x \times L_y \times 2 L_z$ is used; twice the vertical domain size is required for the use of Green's function in solving the potential 
with vacuum boundaries.   Finally, the gravitational potential is mapped back to the original time using the third-order conservative remap that was used on the particle mass density.

The resulting $\Phi_{\rm p}$ is a cell-centered quantity, and we calculate the forces at the cell center using a central finite difference method over three grid cells in every direction.  Thus, the force (Equation~\ref{grav_force}) in the $x$-direction in cell $(i, j, k)$ is calculated by

\begin{equation}
\label{force_finite_difference}
F^{i,j,k}_{x,g} = -\frac{\Phi^{i+1,j,k}_{\rm p} - \Phi^{i-1,j,k}_{\rm p}}{2\Delta x},
\end{equation}

\noindent
and similarly for the $y$ and $z$ directions.  The forces then have to be interpolated back to the location of the particles, and we again use the TSC method.  We add the self-gravity force to the particles simultaneously with the drag force.

The boundary conditions are the shearing-periodic boundaries in the radial direction, purely periodic in the azimuthal direction, and a modified outflow condition in the vertical direction in which the
gas density is extrapolated via an exponential function into the grid zones \cite[][Li et al. 2016 in prep]{simon11a}. This latter boundary condition is
not standard in shearing box setups for the streaming instability as the small domain size makes it difficult to prevent substantial outflow in the vertical direction.  However, in a forthcoming publication by the authors (Li et al. 2016 in prep), we have experimented with the vertical boundary conditions and find that coupled with a routine to renormalize the total mass in the domain to make it constant in time, the outflow boundaries work reasonably well.  
  
The boundary conditions for the gravitational potential are essentially the same as the hydrodynamic variables; shearing-periodic in $x$ and purely periodic in $y$.  The vertical boundary conditions are open, and the
potential is calculated in the ghost zones via a third order extrapolation.

\subsection{Particle Self-Gravity Tests}

In this section, we carry out two tests of our numerical algorithm for particle self-gravity. The first test is the collapse of a uniform density sphere of particles.  The second test
consists of the self-gravitating, shearing wave of particles described in the Supplementary Material of \cite{johansen07a}.

\subsubsection{Spherical Collapse}

The collapse of a uniform density sphere under its own gravity has a simple analytic solution with which to compare the numerical solution. The equation of motion for a test particle at radius $r(t)$ starting at the outer edge of the sphere is 

\begin{equation}
\label{sphere_motion}
\frac{d^2 r}{dt^2} = - \frac{G M}{r^2}
\end{equation}

\noindent
where $M$ is the total mass within the sphere and remains constant in time.  Parameterizing the time dependence of $r(t)$ via $\alpha(t)$, we assume that $r(t) = r_0 {\rm cos}^2\left(\alpha(t)\right)$, where
$r_0$ is the initial radius of the uniform density sphere, and find that

\begin{equation}
\label{alphat}
\alpha + \frac{1}{2} {\rm sin}\left(2\alpha\right) = \sqrt{\frac{2 G M}{r_0^3}} t
\end{equation}

The analytical solutions to the radius and mass density are given by 

\begin{equation}
r(t)/r_0 = {\rm cos}^2\alpha
\end{equation}

\begin{equation}
\rho_{\rm p}(t)/\rho_{{\rm p},0} = \frac{1}{{\rm cos}^6\alpha}
\end{equation} 

\noindent
where $\alpha$ is given by Equation~(\ref{alphat}).

To test our self-gravity solver against this analytic solution, we initialize a sphere of particles with radius $r_0 = 0.25$ on a cubic domain of $L_x \times L_y \times L_z = 1 \times 1 \times 1$ resolved by $96^3$ zones. Every zone within the radius of the sphere has one particle per grid cell initially.  We turn off gas drag on the particles in order to only test the particle self-gravity module. We use TSC interpolation, and the semi-implicit particle integrator. The boundary conditions are open in all three dimensions, and consequently we use the Green's function approach of \cite{koyama09} to solving for the potential with vacuum boundaries; here applied to all three dimensions instead of only the vertical as described above.  We arbitrarily set the value of G to $8\times10^{-3}$.

\begin{figure}[t]
\begin{center}
\includegraphics[width=0.5\textwidth,angle=0]{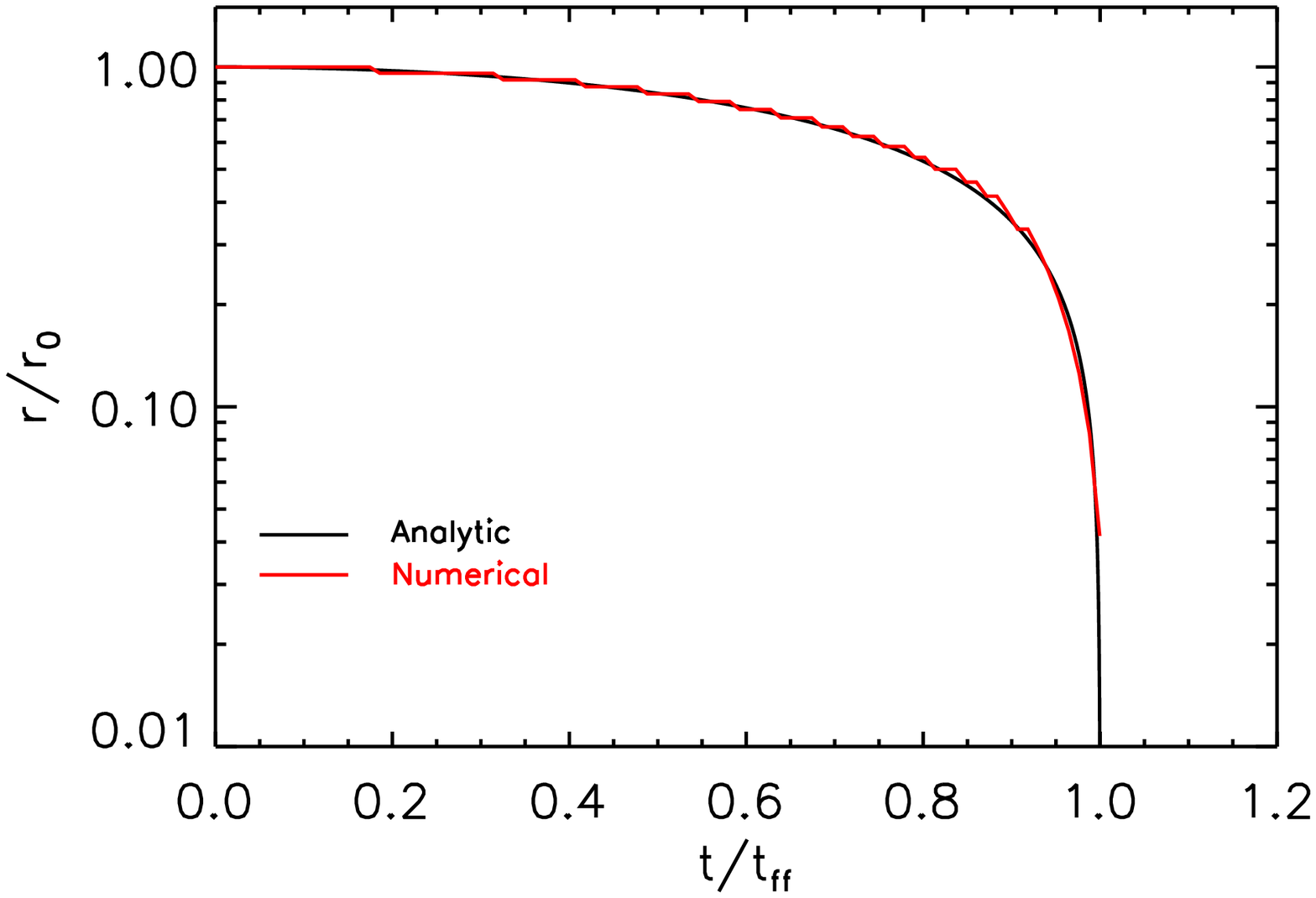}
\includegraphics[width=0.5\textwidth,angle=0]{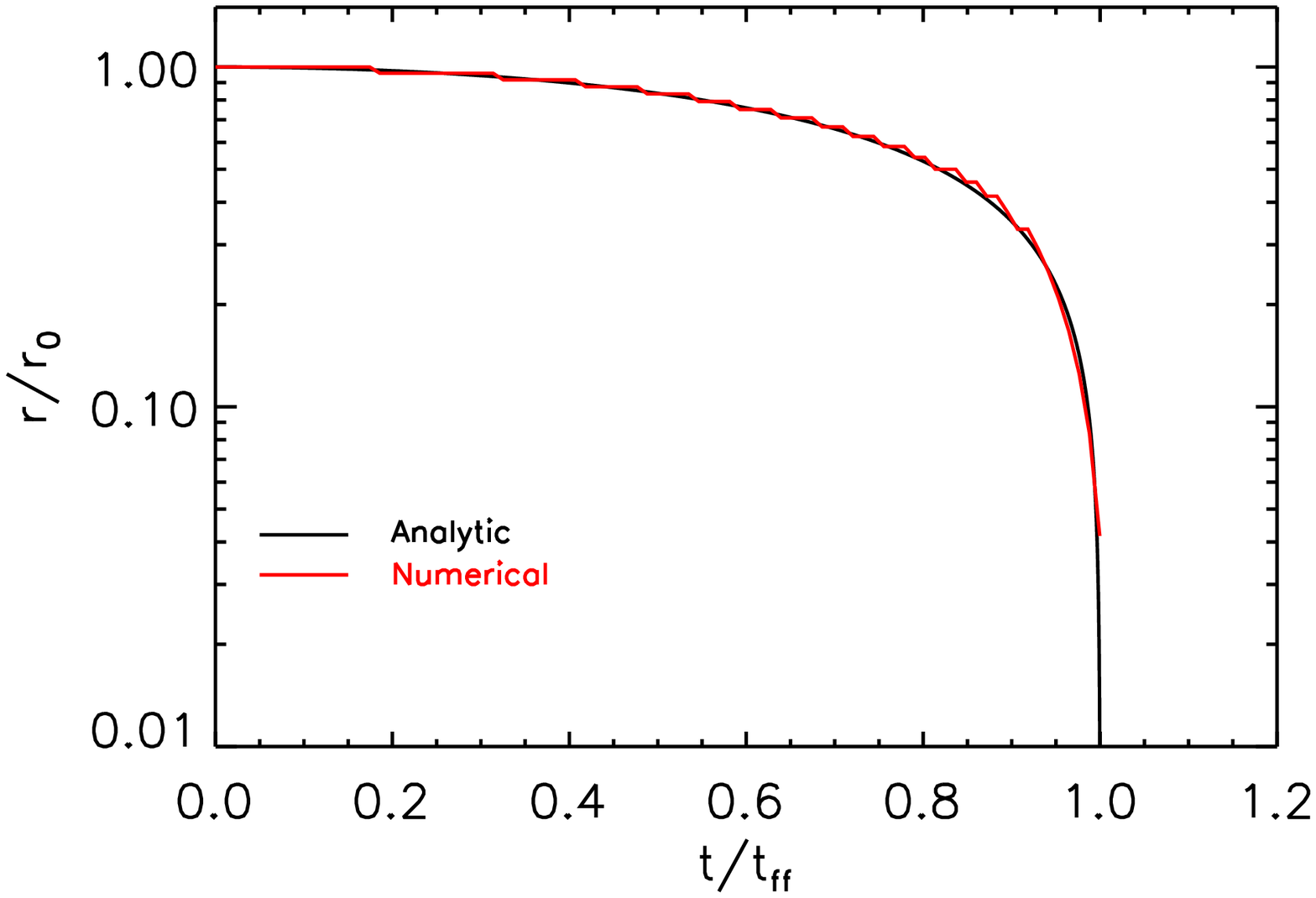}
\end{center}
\caption{Numerical solution to the particle sphere test problem compared to the analytic solution.  The radius of the sphere is shown in the top panel and the particle mass density is shown on the bottom.  In both plots,
the black curves are the analytic solution, and the red curves are the numerical solution.  We plot both the density at the center of the sphere (solid red line) and averaged over the sphere (dashed red line) in the lower plot. Our algorithm for calculating the radial extent of the sphere finds the first cell inward from the boundary in which the particle mass density jumps above 50\% of the density at the sphere's center.  Thus, our algorithm tends to integrate to just slightly larger than the boundary of the sphere, making both the average density and the sphere radius slightly different than the analytic solution.  Otherwise, the numerical solution agrees quite well with the analytic solution.  There is some deviation at late times, at which point we expect the numerical solution to deviate from the analytic due to grid-scale effects.}
\label{collapse}
\end{figure}
 
The solution to this test problem is shown in Fig.~\ref{collapse}, which depicts the evolution of $r(t)/r_0$ and $\rho_{\rm p}(t)/\rho_{{\rm p},0}$ versus time in units of the free fall time $t_{\rm ff} \equiv \sqrt{3\pi/32 G \rho_{{\rm p},0}}$ where $\rho_{{\rm p},0}$ is the initial particle mass density.  
As the figure shows, the numerical solution matches the analytic solution quite well. 
The differences between the two curves that show up at late times is a result of the FFT solver becoming less accurate as the particles are concentrated into fewer and fewer grid cells.  At such particle densities, the
discretization error of the FFT solution starts to dominate, as we discuss more below.  Even with these errors, the numerical solution does not deviate too much from the analytic solution.  This limitation should be kept in
mind when considering the formation of planetesimals.  In fact, the natural softening length of our FFT solver is $\sim \Delta x$ where $\Delta x$ is the length of a grid cell.\footnote{The grid cell is uniform along all three dimensions; $\Delta x = \Delta y = \Delta z$.} 

To further test our gravity module, we have also implemented a direct summation method for calculating the mutual gravitational forces between particles. This method is exact, but scales as $\mathcal{O}(N^2)$, where $N$ is the number of particles.  For large numbers of particles the direct summation method becomes prohibitively expensive.  However, for small numbers of particles we can use this method to compare the forces calculated by the FFT method with the forces from the direct summation, which will be exact to within machine precision. The comparison also tests the accuracy of our interpolation scheme since the direct sum method doesn't require any interpolation of the gravitational forces to the particle locations.  We calculate the error in the force in the $i$-th direction as

\begin{equation}
\label{error}
\epsilon_i = \frac{1}{N_{\rm par}} \sum\limits_{n=1}^{N_{\rm par}}\frac{F_{n,i,{\rm FFT}} - F_{n,i,{\rm direct}}}{|F_{n,i,{\rm direct}}|}
\end{equation}

\noindent
where $F_{n,i,{\rm FFT}}$ is the force due to self-gravity on particle $n$ calculated via the FFT method and $F_{n,i,{\rm direct}}$ is the same but calculated with the direct summation method.  We run the particle sphere collapse problem at a resolution of $32^3$ (and again, one particle per grid cell within the sphere initially) and find that the errors are typically very small early on (on the order of $10^{-5}$ or less).  As time progresses, the errors grow; as the particles get more concentrated within a smaller number of grid cells, the truncation level error from the particle self-gravity solver will become more and more dominant. At $t = 0.5t_{\rm ff}$, the average errors are approximately $1$\% or less, and near $t \approx t_{\rm ff}$, the average errors are on order of a few percent.

\subsubsection{Self-Gravitating, Shearing Wave}

While the spherical collapse problem tests the core of our Poisson solver, we require a problem to test the implementation of this solver in the shearing box setup.  To this end, we employ the linearized self-gravitating, shearing particle wave with gas drag described in detail in Section 1.3.1 of the Supplementary Material of \cite{johansen07a}.  Following that work, we linearize the continuity equation, equation of motion, and Poisson's equation for a particle ``fluid" of density $\rho_p = \rho_{p,0} + \rho_p'$, velocity ${\bmath v} = {\bmath v_0} + {\bmath v'}$, and self-gravity potential $\Phi = \Phi_0 + \Phi'$ where prime represents the linear perturbation on the background.  The perturbation is assumed to be of the form $q'(t,x,y) = \delta q(t) {\rm Exp}[{\rm i}(k_x(t) x+ k_y y)]$.  Note that we assume uniformity along the vertical direction.  The evolution of the density and velocity perturbations are then governed by the following equations,

\begin{equation}
\label{perturbed_continuity}
\frac{{\rm d} \delta \rho_p}{{\rm d} t} = -\rho_{p,0} {\rm i}\left[k_x(t)\delta v_x + k_y \delta v_y\right],
\end{equation}

\begin{equation}
\label{perturbed_vx}
\frac{{\rm d} \delta v_x}{{\rm d} t} = 2 \Omega \delta v_y + \frac{4\pi {\rm i} G k_x(t) \delta \rho_p}{k_x^2(t) + k_y^2} - \frac{\delta v_x}{t_{\rm stop}},
\end{equation}

\begin{equation}
\label{perturbed_vy}
\frac{{\rm d} \delta v_y}{{\rm d} t} = -\frac{1}{2} \Omega \delta v_x + \frac{4\pi {\rm i} G k_y \delta \rho_p}{k_x^2(t) + k_y^2} - \frac{\delta v_y}{t_{\rm stop}},
\end{equation}

\noindent
where the radial wavenumber $k_x$ is time-dependent,

\begin{equation}
\label{kx_eqn}
k_x(t) = k_x(0) + q \Omega t k_y.
\end{equation}

We set $G = t_{\rm stop} = \Omega = \rho_{p,0} = 1$, ${\bmath v_0} = \Phi_0 = \delta \Phi = \delta v_x = \delta v_y = k_x(0) = 0$, $k_y = 1$, $q = 1.5$, and $\delta \rho_p = 10^{-4}$.  We solve Equations~(\ref{perturbed_continuity})-(\ref{perturbed_vy}) using a simple finite difference algorithm to obtain our linear solution to the self-gravitating, shearing wave problem in the linear limit.  For the numerical test of our algorithm, we initialize a grid of size
 $L_x \times L_y \times L_z =  2\pi \times 2\pi \times 0.2$ at resolution $128\times128\times6$ and with one particle per cell initially; the small size/resolution in the vertical direction makes the problem effectively two-dimensional while still testing the full 3D FFT algorithm. In this setup, the boundary conditions are shearing-periodic in $x$ and purely periodic in both $y$ and $z$.  Note that we turn off particle feedback to the gas for this test problem. 

\begin{figure}[t]
\begin{center}
\includegraphics[width=0.5\textwidth,angle=0]{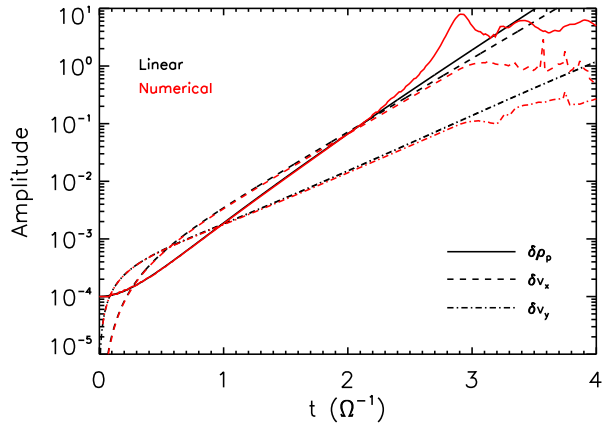}
\end{center}
\caption{Amplitude of the perturbations versus time (in units of $\Omega^{-1}$) in the self-gravitating, shearing wave test.  The red lines show the numerical solution, whereas the black lines show the linear solution.  The solid lines
are the density perturbation, $\delta \rho_p$, the dashed lines are the $x$ velocity perturbations, $\delta v_x$, and the dot-dashed lines are the $y$ velocity perturbations, $\delta v_y$. There is excellent
agreement between the numerical and linear solutions up to amplitudes of $\sim 10^{-1}$, where the numerical solution begins to go non-linear.  }
\label{shear_test}
\end{figure}

The time evolution of the amplitude of the perturbations are shown in Fig.~\ref{shear_test}.  The numerical solution agrees very well with the linear solution up to amplitudes of $\sim 10^{-1}$.  At this point, the
numerical solution approaches the non-linear regime, and agreement between the linear and numerical solutions is not expected.  This test problem, which includes several essential ingredients relevant
to the streaming instability calculations in this paper, (i.e., shear, self-gravity, and drag), demonstrates the validity of our self-gravity implementation in the shearing box setup.

\subsection{Simulation setup}
\label{setup}

All of our streaming instability simulations use a shearing box domain of size $L_x\times L_y\times L_z$ = $0.2H\times0.2H\times0.2H$. Here,
 $H$ is the vertical scale height of the gas, which is initialized with a hydrostatic (Gaussian) vertical profile,
 
 \begin{equation}
 \label{gas_profile}
 \rho_g = \rho_0 {\rm exp}\left(\frac{-z^2}{2H^2}\right),
 \end{equation} 
 
 \noindent
 where $\rho_0$ is the mid-plane gas density. We choose code units so that the standard gas parameters are unity; $\rho_0 = H = \Omega = \cs = 1$. Given the limited vertical extent
 of our domain, the gas density does not vary significantly. 
 
The particles are distributed uniformly in $x$ and $y$ and
with a Gaussian profile in $z$.  The scale height of this profile is $H_p = 0.02H$.  While we
add random noise to the particle locations to seed the streaming instability, it is worth mentioning that the particles
are not initially in an equilibrium state. There is radial motion induced by
the radial drift term $\eta v_{\rm K}$ as described above, and the lack of a pressure
gradient to counteract vertical gravity means that the particles will settle towards the disk mid-plane.  

Four dimensionless quantities characterize the evolution of the streaming instability and the ability of dense clumps to gravitationally collapse. The streaming instability depends upon the dimensionless stopping time,
\begin{equation}
 \tau \equiv t_{\rm stop} \Omega^{-1},
\end{equation}
which characterizes the aerodynamic interaction between a single particle species and the gas, the 
metallicity,
\begin{equation}
 Z = \frac{\Sigma_p}{\Sigma_g}
\end{equation}  
which is the ratio of the particle mass surface density $\Sigma_p$ to the gas surface density $\Sigma_g$, and 
a radial pressure gradient parameter that accounts for the sub-Keplerian gas in real disks
\begin{equation}
\label{headwind}
\Pi \equiv \eta v_{\rm K}/\cs.
\end{equation}
This radial pressure gradient produces a headwind on the particles, which
has velocity a fraction $\eta$ of the Keplerian velocity.
For all of our runs, $\tau = 0.3$, metallicity 
is moderately super-Solar, $Z = 0.02$, and $\Pi = 0.05$.

With the inclusion of self-gravity, an additional parameter is needed to describe the relative importance of 
self-gravity and tidal shear. We
define a parameter $\tilde{G}$,
\begin{equation}
\label{Gtilde}
\tilde{G} \equiv \frac{4\pi G\rho_0}{\Omega^2},
\end{equation}
which describes the strength of self-gravity in the simulation.  Physically, varying $\tilde{G}$ is equivalent to 
changing the gas density (and thus through assuming the same $Z$, the particle mass density) or
the strength of tidal stretching (i.e., changing $\Omega$) or both. Thus, in a real disk, $\tilde{G}$ will
vary with radius, and in a minimum mass solar nebula model \cite[MMSN;][]{hayashi81}, this parameter increases
very gradually with radius.

In what follows, we convert code units to physical units
using the mass unit $M_0 = \rho_0 H^3$ and assuming a radius of 3 AU in a MMSN. For $\tilde{G} = 0.05$,
this equates to $M_0 =  6.5\times10^{26} {\rm g} \approx 720 M_{\rm Ceres}$.  This physical unit conversion depends on $\tilde{G}$, as
we discuss further below. 

Our fiducial simulation is a relatively low resolution run with $N_x\times N_y\times N_z$ = $128\times128\times128$ gas 
zones and 2,400,000 particles. For this simulation, we turn on self-gravity at a time $t_{\rm sg} = 170\Omega^{-1}$ and 
set $\tilde{G} = 0.05$.

\subsection{Parameter variation}

We then explore parameter space by varying one parameter in each
simulation subset (demarcated in Table~\ref{tbl:sims} by separate boxes).  In the first subset, we vary the numerical resolution and the number of particles.  We
explore a lower resolution $64\times64\times64$ with 300,000 particles, and two higher resolutions: $256\times256\times256$ with 19,200,000 particles and $512\times512\times512$ with 153,600,000 particles.
In the next subset, we vary the value of $\tilde{G}$ from 0.02 to 0.1; this was chosen to match the test carried out by \cite{johansen12} in which the strength of gravity was varied. 

Finally, we determine the effect of the state at which self-gravity is initiated by examining two diagnostics in a simulation identical to the fiducial one, but with no self-gravity. The first is the maximum value of the volume-averaged particle mass density, $\rho_p$, and the second is a weighted version of the volume-averaged particle mass density, defined as $\sqrt{\langle\rho_p^2\rangle}/\langle\rho_p\rangle$.  In general, these two quantities behave similarly, but while the maximum
density has been used more frequently in the literature to track the degree of clumping via the streaming instability, it can be largely affected by a small number of grid cells (or even one cell), and the latter quantity is more representative  
of the degree of clumping over the entire domain.  Using both of these diagnostics in concert, we chose restart times of $t_{\rm sg} = 40\Omega^{-1}$ (the ``low clumping" case), $t_{\rm sg} = 170\Omega^{-1}$ (medium clumping; this is the same as the fiducial simulation), and $t_{\rm sg} = 240\Omega^{-1}$ (high clumping).  We also run one case with self-gravity turned on from the beginning state $t_{\rm sg} = 0$.   

All of these runs are labelled by the numerical resolution employed, the strength of self-gravity, and any other modifying information.  For example, run SI128-G0.05-low\_clump has 128 zones per dimension, $\tilde{G} = 0.05$,
and is initiated from a ``low clumping" state, as described above.  All simulations are shown in Table~\ref{tbl:sims} with the parameters outlined and any additional, relevant comments.

\subsection{Diagnostics}

There are several diagnostics employed in Section~\ref{results} that we now describe. First is the maximum particle mass density, as described above, as a proxy for the degree
of clumping during the streaming instability.  Another often employed quantity is the particle mass surface density, which is simply the integral of $\rho_p$ over the vertical extent of the domain. 

Most of our diagnostics require that we locate gravitationally bound mass clumps (after self-gravity has been included) and calculate their mass.  Following the basic arguments in \cite{johansen11a}, we calculate $\Sigma_p$
at a given time and use a peak-finding algorithm to locate local maxima in the surface density above a certain threshold.  The mass of any given clump is determined by calculating a circular region surrounding the clump
and iteratively increasing the radius of this region until the mass enclosed within it equals the mass of a clump that is bound by self-gravity compared to tidal effects. Mathematically, we
seek a radius $R$ such that the tidal term, $3\Omega^2R$, and the gravitational acceleration of a test particle at the edge of the clump's Hill sphere, $GM_p/R^2$ are in balance;

\begin{equation}
\label{mass_bound}
M_p = \frac{3\Omega^2 R^3}{G};
\end{equation}

\noindent
when the mass enclosed within our test circle equals $M_p$, we set $R$ as the Hill radius and $M_p$ as the clump mass.

There are two limitations to our algorithm. First, it has difficulty determining the masses of clumps when two or more clumps are very close together in the $xy$ plane.  This is alleviated somewhat by setting a minimum distance, 
equivalent to the Hill radius of a 0.1 $M_{\rm Ceres}$ mass (which, as we will see, is consistent with masses at the higher end of the resulting mass distributions), below which the algorithm counts two clumps as being one.  Furthermore, we subtract contributions of neighboring clumps that are beyond this minimum distance but still within each clump's Hill radius.  This limitation with overlapping clumps is particularly troublesome at early times when gravitationally bound clumps have yet to fully form, yet there are still peaks in $\Sigma_p$.  
The error associated with overlapping clumps commonly manifests itself as very sharp peaks or troughs in the time evolution of the clump masses.  While this limitation causes issues on short timescales when clumps temporarily interact, it does not affect the clump mass on long timescales.  As a first order approach to remove this effect, we apply a simple boxcar smoothing technique to the time evolution of clump masses in all that follows.

The second limitation arises from applying a two-dimensional approach to a fully three-dimensional problem. However, even though clumps exist in three dimensions, the particle layer is vertically thin enough to make the two-dimensional clump finding approach a good approximation. Indeed, the algorithm appears to do a sufficient job at finding clumps that remain gravitationally bound as the simulation progresses and does not find too many false positives.  Furthermore, as shown in Section~\ref{results}, the diagnostics that employ this algorithm return results roughly consistent with \cite{johansen12}, lending support to the idea that our analysis works reasonably well.

From this clump finding algorithm, we calculate both the time history of the mass of the formed clumps and the mass distribution function for these clumps. Many of the simulations form only small numbers of clumps, so we seek a 
simple functional form to fit to the mass distribution. Consistent with prior work, a power-law  
differential mass distribution,

\begin{equation}
\label{mass_dist}
\frac{{\rm d}N}{{\rm d}M_p} = C_1 M_p^{-p},
\end{equation}

\noindent
works well.  Here, $C_1$ is a constant that for the purposes of this paper is arbitrary. We also consider the 
cumulative mass distribution,

\begin{equation}
\label{cum_dist}
N(>M_p) = C_2 M^{-c}, 
\end{equation}

\noindent
where $C_2$ is another arbitrary constant.  We can easily translate Equation~(\ref{mass_dist}) to a size distribution to find ${\rm d}N/{\rm d}r \propto r^{-q}$ where $q = 3p-2$.

Under the assumption that the data is drawn from a power-law distribution, the power law index $p$ of the 
differential distribution can be determined directly from the set of measured masses $M_{p,i}$ using a 
maximum likelihood estimator \cite[MLE;][]{clauset09}. From Equation~(3.1) in \cite{clauset09}, 
we estimate the value of the power law index by 

\begin{equation}
\label{mle_p}
p = 1 + n \left[\sum^n_{i=1} {\rm ln}\left(\frac{M_{p,i}}{M_{p,{\rm min}}}\right)\right]^{-1},
\end{equation}

\noindent
where $M_{p,{\rm min}}$ is the minimum value of the planetesimal mass in our data, and $n$ is the total number of planetesimals.  The
error is given by their Equation (3.2),

\begin{equation}
\label{mle_err}
\sigma = \frac{p-1}{\sqrt{n}}.
\end{equation}
Using the MLE for $p$ avoids any binning step, which can introduce bias into the estimate. To visually 
represent the differential distribution, however, we make a local estimate of ${\rm d}N / {\rm d}M_p$ 
by taking ${\rm d}M_p$ to be half the distance in the mass co-ordinate between a given planetesimal 
and its nearest neighbors. A least squares linear fit in log-log space to ${\rm d}N/{\rm d}M_p$ vs $M_p$ gives an alternate measure of $p$, that is very roughly consistent 
but generally slightly smaller.

Irrespective of the statistical methods used, two cautions are in order. First, the underlying 
distribution of planetesimal masses formed in the simulations cannot in reality be a simple 
power-law, because there is zero probability of forming a body with a mass greater than the 
total mass of solids in the domain. If the actual distribution is a truncated power-law, then 
our naive fit will over-estimate the value of $p$. Second, our lower 
resolution runs form rather small number of planetesimals, and any method for estimating 
$p$ in this situation is noisy, possibly biased, and liable to return a non-gaussian distribution 
of slope estimates. Simple experiments suggest that only the higher resolution runs, for 
which there are on order 100 planetesimals, can be expected to return robust estimates of the distribution.


\section{Results}
\label{results}

In this section, we present the results from our various parameter studies.  We first describe the properties of the fiducial simulation SI128-G0.05 in detail, followed by quantifying the effects of changing the strength of gravity,
the numerical resolution, and the degree of clumping when self-gravity is initiated.  Each study is described in its own subsection and the simulations corresponding to each study
are presented in Table~\ref{tbl:sims}.


\begin{figure*}[ht]
\begin{center}
\includegraphics[width=0.8\textwidth,angle=0]{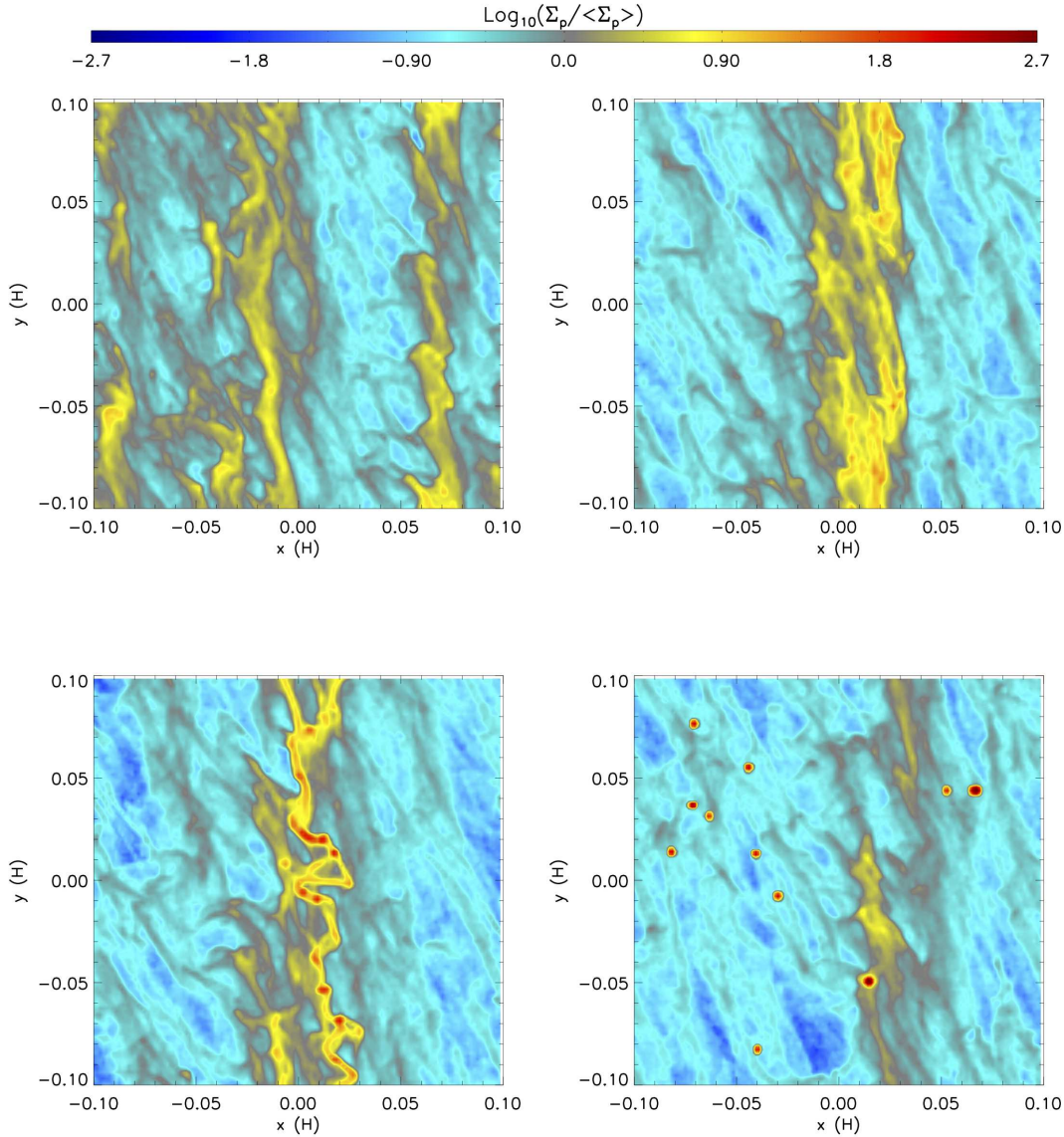}
\end{center}
\caption{Four snapshots during the fiducial simulation, SI128-G0.05.  Shown is the logarithm of the vertically integrated particle surface density normalized to the average particle surface density.  Time increases from top left
to bottom right.  The bottom left is shortly after self-gravity is turned on, and the bottom right is well after the planetesimals have formed.}
\label{snapshots_fiducial}
\end{figure*}

\subsection{Fiducial Run}
\label{fiducial}

In this section, we describe some properties of our fiducial run, SI128-G0.05.  Self-gravity is switched on at $t = 170\Omega^{-1}$, after
which the mutual gravitational attraction between particles causes the particle density to increase; in units of the initial mid-plane gas density, the maximum particle density rapidly increases to $\sim 5\times10^3$, and then slowly increases to $\sim 3\times10^4$ afterwards. 

\begin{figure}[ht]
\begin{center}
\includegraphics[width=0.5\textwidth,angle=0]{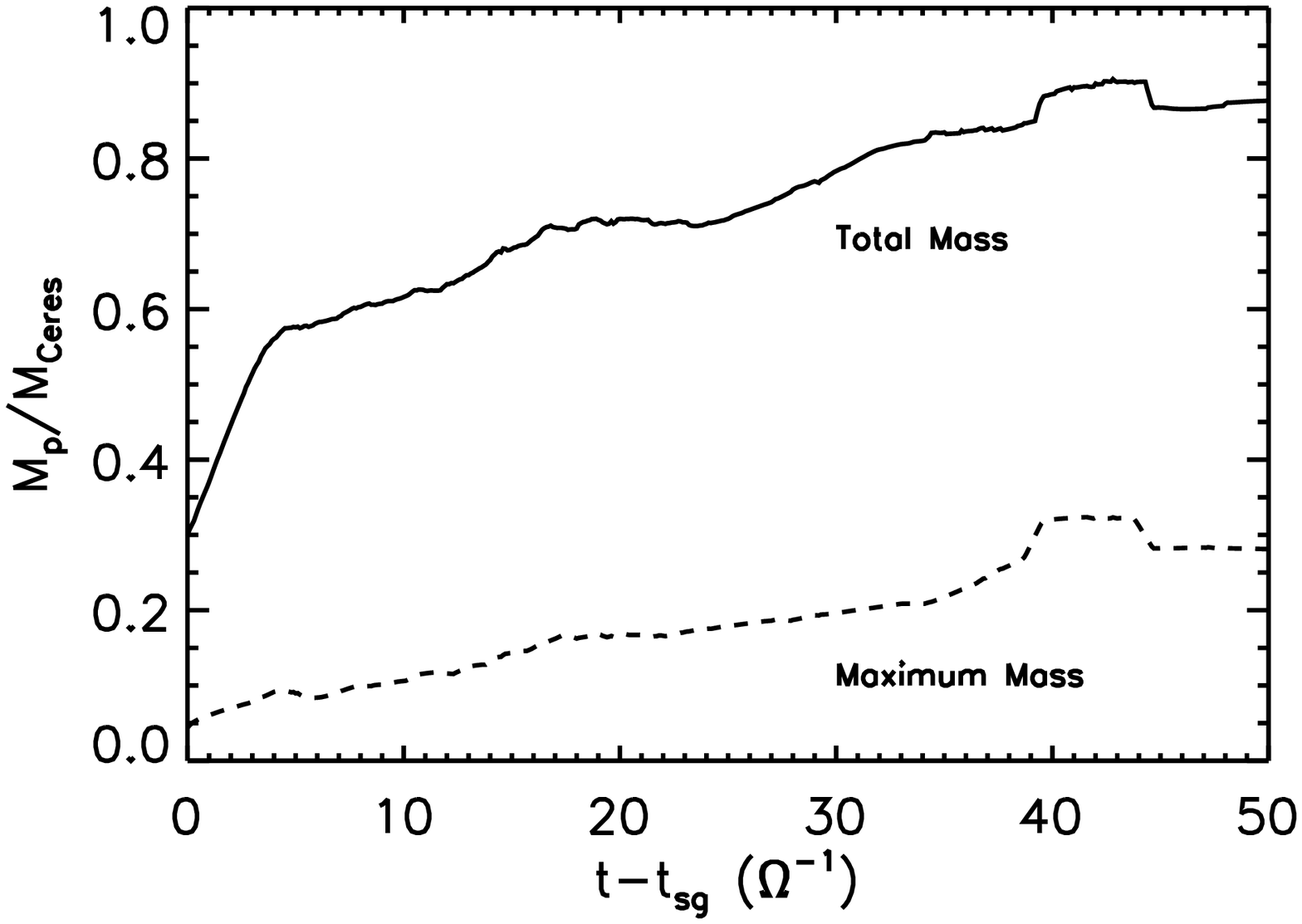}
\end{center}
\caption{Evolution of the total mass in planetesimals (solid line) and maximum planetesimal mass (dashed line) over time in our fiducial simulation.  The mass is given 
in units of the mass of Ceres. Time is in units of $\Omega^{-1}$ and is measured from the initialization of self-gravity onward.  We find general agreement with a similar run in \cite{johansen12}, but there
are some differences in the initial growth rate of planetesimals.}
\label{fiducial_mass}
\end{figure}

A time progression of the particle surface density is shown by a series of snapshots in Fig.~\ref{snapshots_fiducial}.  The streaming instability produces several azimuthally extended structures at early times, which eventually form into one large clumping structure.  After self-gravity turns on, some of the high density regions collapse and become gravitationally bound structures with a variety of masses.   Near the end of the simulation, there are on order 10 of these bound structures, 
to which we refer from now on as planetesimals.

The evolution of the mass of these planetesimals is shown in Fig.~\ref{fiducial_mass}; 
both the total mass and maximum mass are shown.  This evolution can be compared to Fig. 13 of \cite{johansen12}.  Although, most of the simulations in \cite{johansen12} contain collisional microphysics, they find that
the mass evolution of planetesimals only depends very weakly on this, if at all.   We find that the total and maximum masses in units of the mass of Ceres are approximately the same by $t-t_{\rm sg} = 20\Omega^{-1}$ in our fiducial
simulation and the $\tilde{G} = 0.05$ simulation of \cite{johansen12}.  Our simulations show a faster growth rate for the masses of planetesimals.  Given the difficulties associated with
the clump finding analysis early on, some differences are not surprising.  However, that we find approximately the same values for the maximum and total masses is encouraging.

Finally, we analyze the mass and size distribution of these planetesimals. We find that there are 13 clumps, and roughly, the number of these clumps increase towards the low mass end. 

Choosing the point at which to calculate
the mass distribution is a bit subtle. Our general approach is to visually inspect the particle mass density (e.g., Fig.~\ref{snapshots_fiducial}) and choose a time shortly after planetesimals have formed (so as to not sample later times when 
these planetesimals have substantially grown in mass due to accretion of smaller particles and/or mergers) and have become separate objects (i.e., no significant overlap with the over density of mass from which they formed).  
We have calculated the mass distribution at three different times; $t-t_{\rm sg} = 10.9\Omega^{-1} $, $t-t_{\rm sg} = 12.3\Omega^{-1} $ and $t-t_{\rm sg} = 13.6\Omega^{-1}$.  While the large scatter in the differential mass makes a comparison between the different snapshots and a precise quantification of the power law slope difficult, we find approximate ``by-eye" agreement in the values of $M_p$ and ${\rm d}N/{\rm d}M_p$.   To reduce the noise inherent in the
differential mass distributions, we also checked the cumulative distributions at these three times and found excellent agreement. Thus, we are confident that there is not a substantial variation in the clump properties over short timescales.

\begin{figure}[ht]
\begin{center}
\includegraphics[width=0.45\textwidth,angle=0]{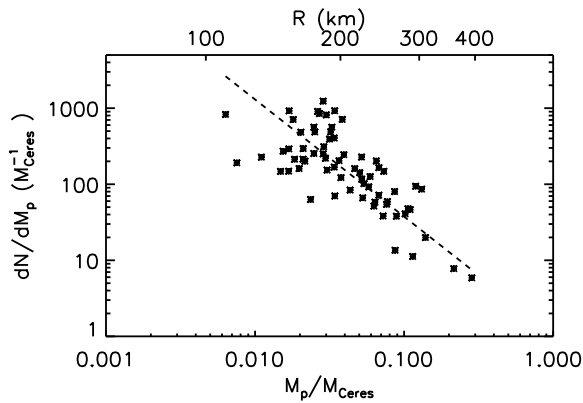}
\end{center}
\caption{Differential mass distribution functions for the fiducial simulation along with
the four simulations that were restarted at different times to improve statistics. 
Both the mass and the differential mass function are given in units of Ceres mass. A best fit power law is over plotted as a dashed line of the corresponding color of the data. 
The best fit power slaw slope is $p = 1.5\pm 0.1$.}
\label{dist1d_fiducial_coadd}
\end{figure}

The significant scatter present in the mass distribution brings into question any robust quantification of the value of $p$.  The best fit value of $p$ from this simulation is $p = 2.0 \pm 0.3$; clearly there is a large error on $p$. In order to improve the statistics, we have run four additional simulations with the same parameters as this fiducial run, but restarted from the ``parent" non-self-gravitating simulation at different times.  Specifically, we initiated self-gravity at $20\Omega^{-1}$ before, $10\Omega^{-1}$ before, $10\Omega^{-1}$ after, and $20\Omega^{-1}$ after the fiducial run. These runs are included in Table~\ref{tbl:sims}; the run appended with ``tm20" (``tp20") means 
self-gravity is initiated at the fiducial case restart time minus (plus) $20\Omega^{-1}$. We chose these relatively large time displacements to reduce the likelihood that we would be sampling planetesimals that occur from quite nearly the same initial conditions; in such a case, the formation of planetesimals would not be independent. The result is shown in Fig.~\ref{dist1d_fiducial_coadd}.  We fit this data with a power law and now find $p = 1.5 \pm 0.1$; the error has been reduced by roughly a factor of three.

There is a question of whether or not the properties of the distribution of planetesimal masses depend on the time at which self-gravity is activated.  We carry out such an investigation below, and as described further in that section, we find that these properties do not appear to have any strong correlation with the initial state,  justifying this approach.


\subsection{Effect of Resolution}

\begin{figure*}[!ht]
\begin{center}
\includegraphics[width=0.8\textwidth,angle=0]{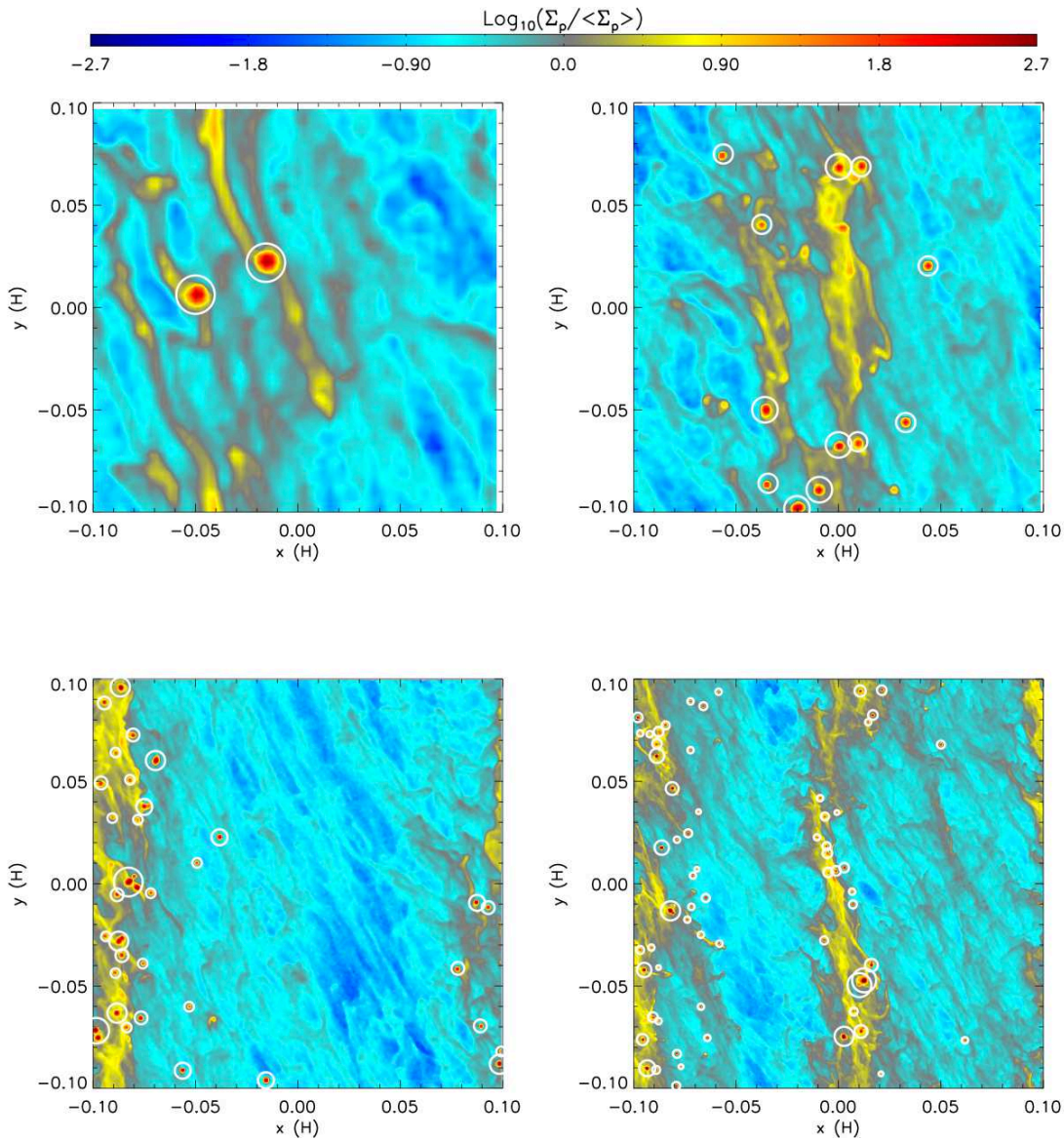}
\end{center}
\caption{Snapshot of planetesimal formation at each of the four different resolutions.  Shown is the logarithm of the vertically integrated particle surface density normalized to the average particle surface density for runs SI64-G0.05 at $t-t_{\rm sg} =11.4\Omega^{-1}$ (top left), SI128-G0.05 at $t-t_{\rm sg} = 12.3\Omega^{-1}$ (top right), SI256-G0.05 at $t-t_{\rm sg} = 18.7\Omega^{-1}$ (bottom left) and SI512-G0.05 at
$t-t_{\rm sg} = 7.6\Omega^{-1}$ (bottom right).  Each planetesimal is marked via a circle of the size of the Hill sphere.  In some cases of extreme overlap between planetesimals, only one circle is drawn. As resolution is increased, more planetesimals are produced, and the number of smaller planetesimals increase.}
\label{snapshots_res}
\end{figure*}

Figure~\ref{snapshots_res} shows a snapshot of the particle mass surface density for each resolution. As with the fiducial simulation, the time chosen in each case is sufficiently long after self-gravity has been turned on for bound planetesimals to form, but sufficiently early so as to avoid the effect of merger events and additional mass accretion onto the formed planetesimals. The times corresponding to these snapshots are also chosen in the calculation of the mass distribution function below. As expected, the size of the largest planetesimals decrease as the grid scale decreases; the minimum radius follows the grid zone size.  Furthermore, as resolution is increased, the number of planetesimals increases and the smallest bound mass decreases.  From resolution $64^3$ to $512^3$, the number of formed planetesimals at this time are 2, 13, 30, and 53, respectively.

\begin{figure*}[ht]
\begin{center}
\includegraphics[width=0.35\textwidth,angle=-90]{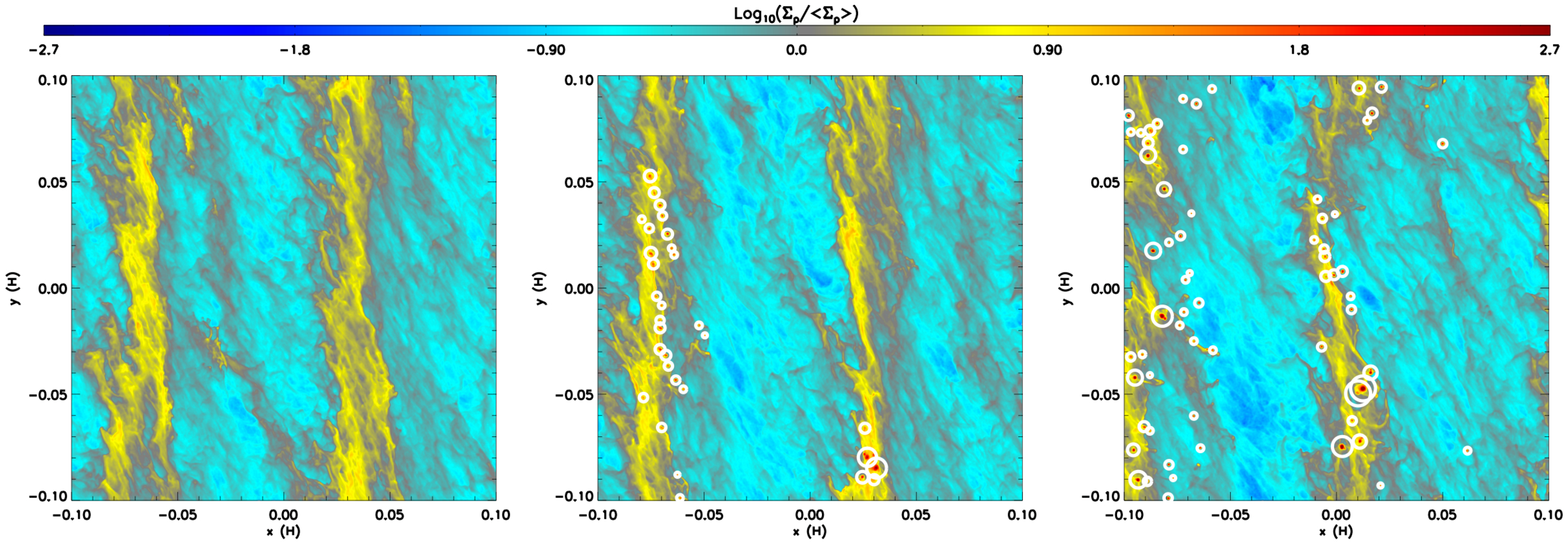}
\end{center}
\caption{Three snapshots during the highest resolution simulation, SI512-G0.05.  Shown is the logarithm of the vertically integrated particle surface density normalized to the average particle surface density.  Time increases from left
to right.  The left panel shows the clumping due to the streaming instability in the absence of self-gravity but right before self-gravity is activated ($t = 110\Omega^{-1}$).  The middle panel corresponds to a point shortly after self-gravity was activated ($t = 112.5\Omega^{-1}$), and the right panel corresponds to a time in which
most of the planetesimals have formed ($t = 117.6\Omega^{-1}$). In the middle and right panel, each planetesimal is marked via a circle of the size of the Hill sphere.  Planetesimals continue to form and to grow in mass during the simulation, as demonstrated by the larger number and increased size of Hill spheres in the right panel.  In some cases of extreme overlap between planetesimals, only one circle is drawn.  Compared with the lower resolution simulation, smaller scale structure and the development of more numerous and smaller planetesimals is observed in the particle density.}
\label{snapshots_highres}
\end{figure*}

In order to show the development of these small scale structures in the highest resolution run, we have plotted the various stages of the streaming instability before and after self-gravity was turned on in Fig.~\ref{snapshots_highres}. As both Fig.~\ref{snapshots_res} and Fig.~\ref{snapshots_highres} show, there is significant structure present on small scales, though the large scale structure remains consistent with the lower resolution runs.  In particular, there are large scale axisymmetric enhancements in the particle mass density.  Once self-gravity has been activated, the smaller structure available at the higher resolution allows smaller mass planetesimals to condense out of the high densities induced by the streaming instability.  Despite the increase in the number of planetesimals at small scales, larger mass planetesimals still form. 

\begin{figure*}[t]
\begin{center}
\includegraphics[width=0.45\textwidth,angle=0]{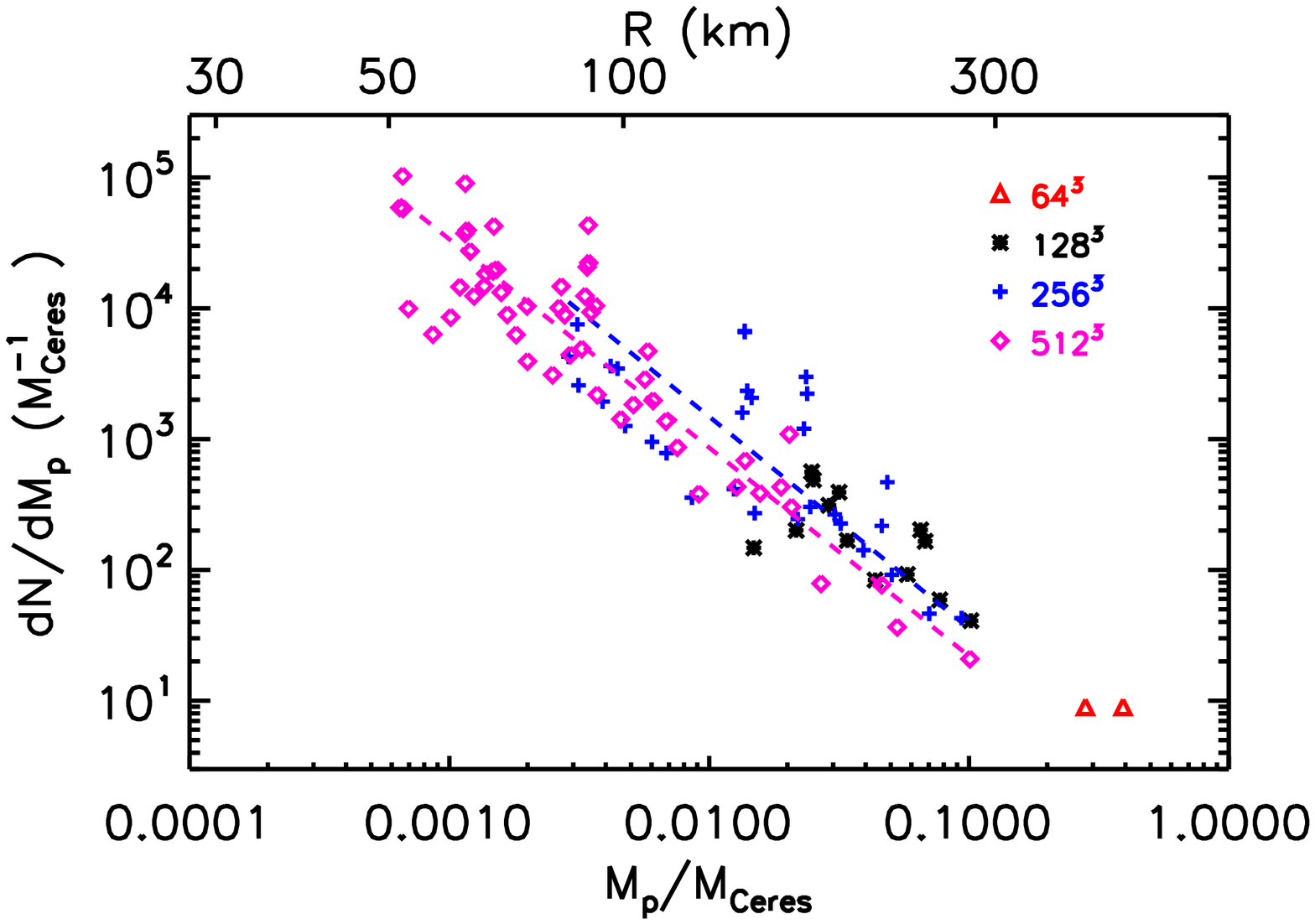}
\includegraphics[width=0.45\textwidth,angle=0]{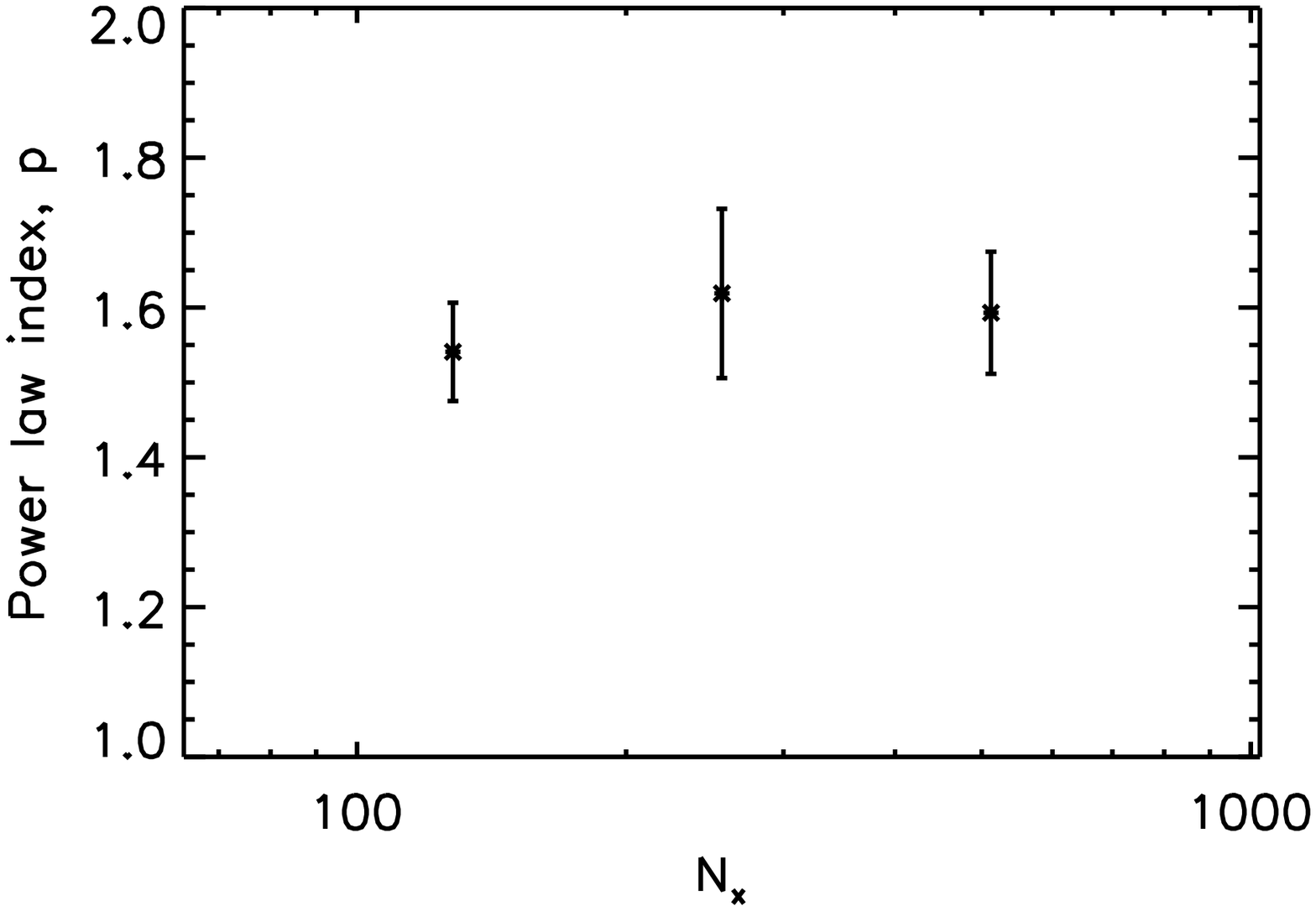}
\end{center}
\caption{Left: Differential mass distribution for SI64-G0.05  at $t-t_{\rm sg} =11.4\Omega^{-1}$ (red triangles), SI128-G0.05 at $t-t_{\rm sg} = 12.3\Omega^{-1}$ (black asterisks), SI256-G0.05 at $t-t_{\rm sg} = 18.7\Omega^{-1}$ (blue crosses) and SI512-G0.05 at $t-t_{\rm sg} = 7.6\Omega^{-1}$ (purple diamonds). Both the mass and the differential mass function are given in units of Ceres mass.  We only include the points from the single fiducial $128^3$ run in order to properly show 
the effect of resolution on the total number of planetesimals. For the two highest resolution simulations, a best fit power law is over plotted as a dashed line of the corresponding color of the data.
Right: Power law index, $p$ versus resolution in number of grid zones along any given dimension with one sigma uncertainties to the fit represented by error bars.  The value of $p$ for $128^3$ here does include the additional four simulations as described in Section~\ref{fiducial}. The power law slope is $p = 1.5 \pm 0.1$ for SI128-G0.05, $p = 1.6 \pm 0.1$ for SI256-G0.05, and $p = 1.6\pm 0.1$  for SI512-G0.05.  With the exception of the lowest resolution, the mass at the high
end of the mass distribution is roughly constant with resolution.  As resolution is increased, there are more low mass planetesimals.  There is no consistent trend of the slope with resolution.}
\label{dist1d_res}
\end{figure*}

This resolution effect is also demonstrated via the differential mass distribution, shown in Fig.~\ref{dist1d_res}.  We do not include a power law fit to either the SI64-G0.05 planetesimals since there are only two objects formed at that resolution or 
to the SI128-G0.05 planetesimals because we only include the single standard run here (in order to more clearly see the effect of the resolution on the number of planetesimals).  Furthermore, the time chosen for the highest resolution run was taken to be the end of the simulation.  It is not entirely clear if planetesimals have stopped forming by this point, and the very high computational expense of this simulation makes integrating further not feasible at this time. 

With the exception of the lowest resolution, the highest mass planetesimals are $M_p \sim 0.1 M_{\rm Ceres}$ for all resolutions. The higher masses (by only a factor of a few) in the $64^3$ run could be related to more rapid growth of the planetesimals that do form since their physical cross section will be larger.  However, since there are only two planetesimals, very small number statistics make this notion
difficult to test.

The power law slope is $p = 1.5 \pm 0.1$ for SI128-G0.05, $p = 1.6 \pm 0.1$ for SI256-G0.05, and $p = 1.6\pm 0.1$  for SI512-G0.05. There is substantial scatter around the best fit power law functions, as the figure shows, though the number of formed planetesimals increases with higher resolution, improving the statistics somewhat.  In the right panel of Fig.~\ref{dist1d_res}, we plot the power law index $p$ with the errors given by Equation~(\ref{mle_err}).  There does not appear to be any significant
trend in the value of $p$ with resolution, and $p \approx 1.4$--$1.8$.

Our best fit value of $p$ for the highest resolution simulation is the same as that found for the $512^3$ simulation of \cite{johansen15}; they found a value of $p = 1.6$ by fitting the cumulative distribution to a power law with an exponential tail.  Furthermore, 
comparing the left panel of Fig.~\ref{dist1d_res} to Fig. 3 of \cite{johansen15} reveals similar scatter about the best fit line for each resolution.


\subsection{Strength of Gravity}
\label{vary_g}

\begin{figure*}[!ht]
\begin{center}
\includegraphics[width=0.45\textwidth,angle=0]{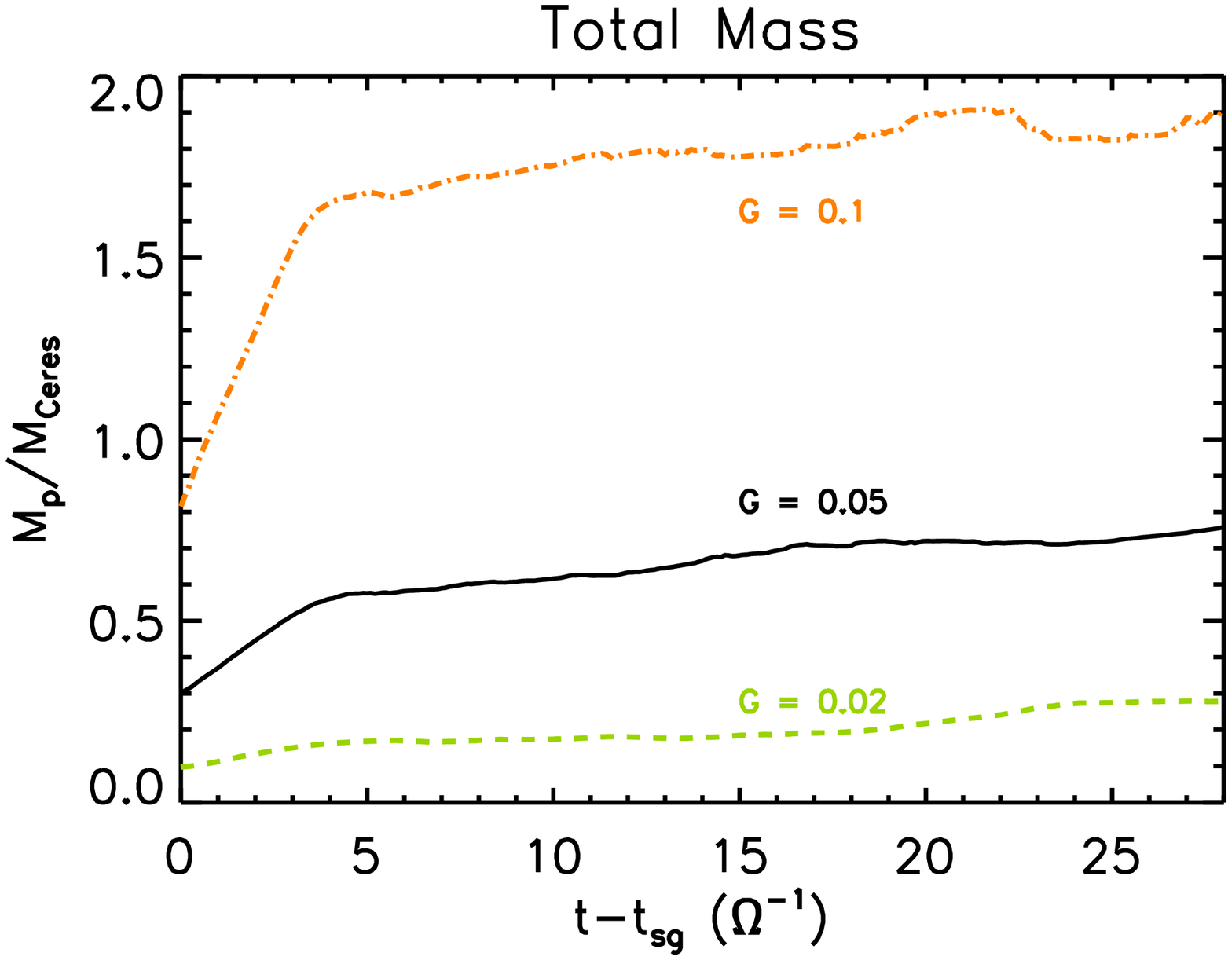}
\includegraphics[width=0.45\textwidth,angle=0]{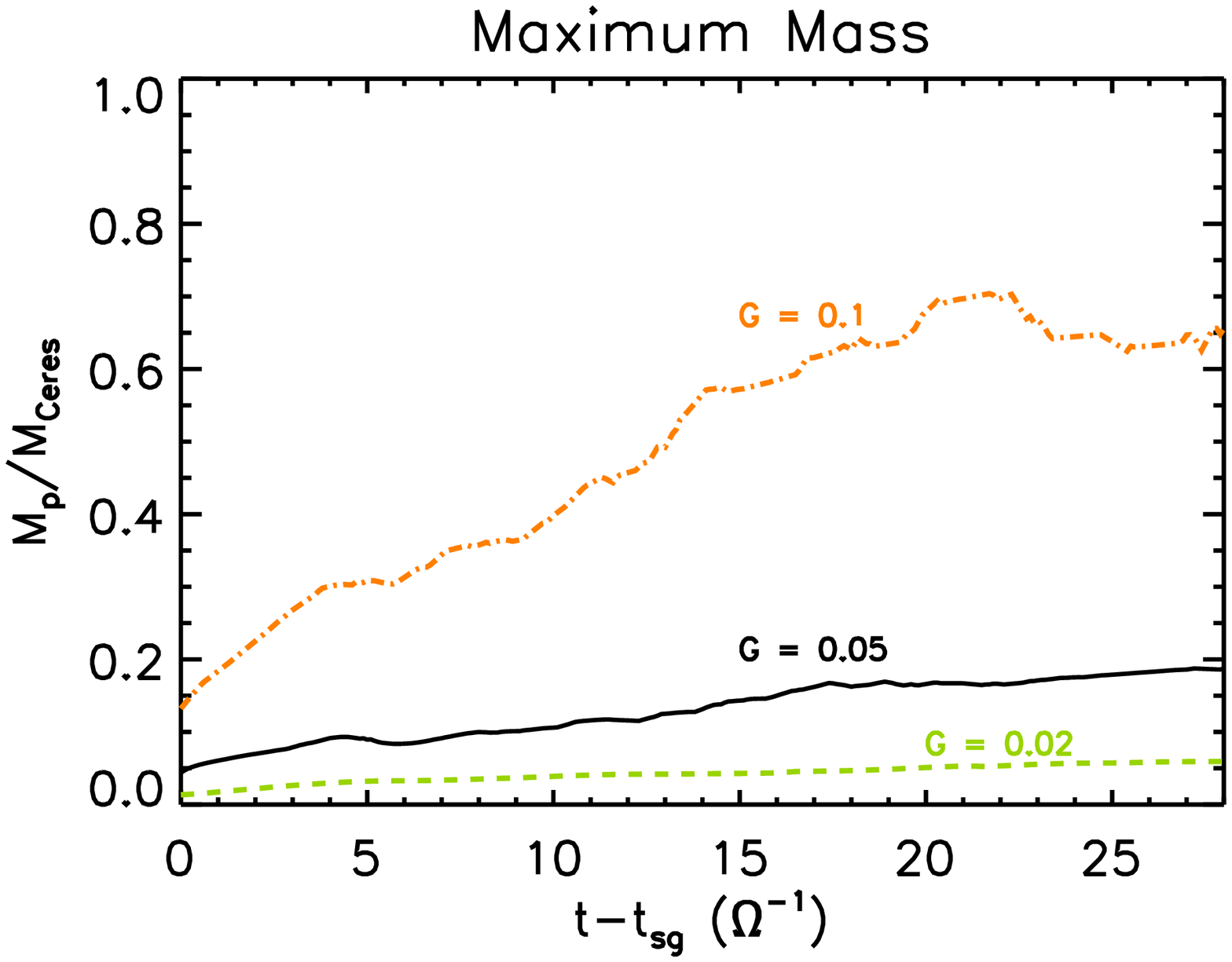}
\end{center}
\caption{Total (left) and maximum (right) mass in planetesimals as a function of time for $\tilde{G} = $ 0.02 (dashed, green line), 0.05 (solid, black line), and 0.1 (dot-dashed, orange line).   The mass is given 
in units of the mass of Ceres, and time is in units of $\Omega^{-1}$ and is measured from the initialization of self-gravity onward.  On the plot, we denote the dimensionless gravity parameter with $G$ instead of $\tilde{G}$. The total and maximum planetesimal masses increase
with increasing $\tilde{G}$. We find general agreement with Fig. 13 of \cite{johansen12}, but there
are some differences in the initial growth rate of planetesimals. }
\label{g_mass}
\end{figure*}

We vary $\tilde{G}$ to match a similar exploration of this parameter in \cite{johansen12}; specifically, $\tilde{G} = 0.02, 0.05, 0.1$.  Note that with the exception of one of the simulations with $\tilde{G} = 0.1$, the
\cite{johansen12} simulations included collision microphysics.  However, as discussed above, these authors concluded that the collisions did not have a strong impact on the formation of planetesimals, making
a comparison between our work and theirs viable.

Figure~\ref{g_mass} shows the mass evolution of the planetesimals for the three values of $\tilde{G}$.  Both the total and maximum planetesimal mass increase with increasing gravity.  This figure can be compared
to Figure 13 of \cite{johansen12}. In general, we find good agreement with their results; the mass values differ by a factor of less than two.  Furthermore, there is no systematic direction in which
the masses differ; i.e., some of the mass values in \cite{johansen12} are larger than what we find for the same $\tilde{G}$ and others are smaller. Given the differences likely present in the clump finding
algorithms, we believe that the agreement is pretty strong.

\begin{figure*}[!ht]
\begin{center}
\includegraphics[width=0.45\textwidth,angle=0]{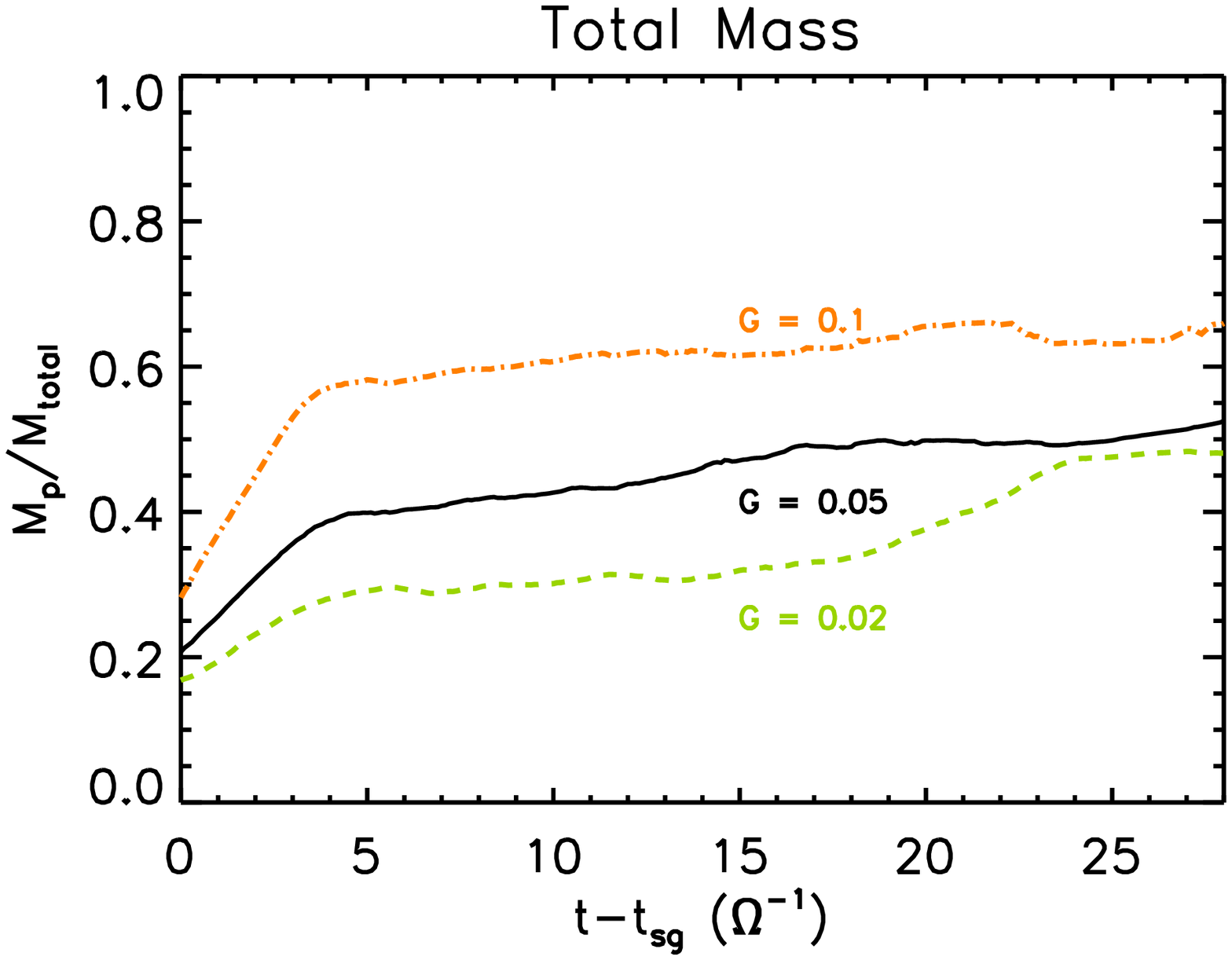}
\includegraphics[width=0.45\textwidth,angle=0]{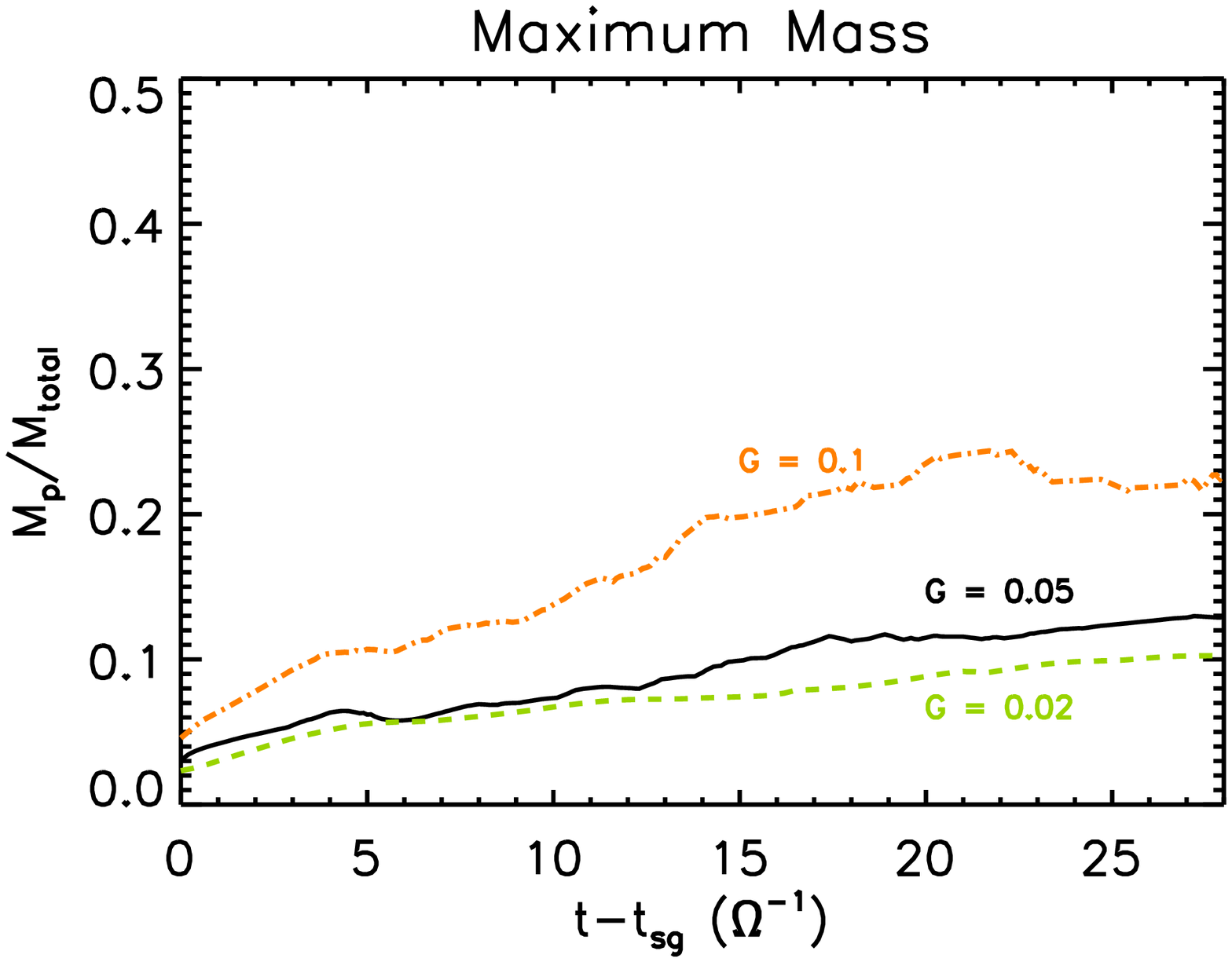}
\end{center}
\caption{Same as Fig.~\ref{g_mass} but with the planetesimal mass normalized to the total particle mass in the domain.  The renormalization of mass reduces the difference between the different $\tilde{G}$ values.  
However, the total and maximum planetesimal masses still increase with increasing $\tilde{G}$.}
\label{g_mass_normalized}
\end{figure*}

In converting the planetesimal mass to physical mass units (to normalize by $M_{\rm Ceres}$), the gravity parameter $\tilde{G}$ is folded into the calculation. Thus, a more meaningful comparison
between these simulations is to renormalize the planetesimal mass to the total mass in the grid, which is {\it independent} of $\tilde{G}$.  This is shown in Fig.~\ref{g_mass_normalized}.
While the total and maximum planetesimal masses increase with $\tilde{G}$ as was the case in Fig.~\ref{g_mass}, the differences between the curves are significantly reduced, and
the maximum mass of SI128-G0.02 and SI128-G0.05 appear to be roughly the same. 

\begin{figure*}[!ht]
\begin{center}
\includegraphics[width=0.45\textwidth,angle=0]{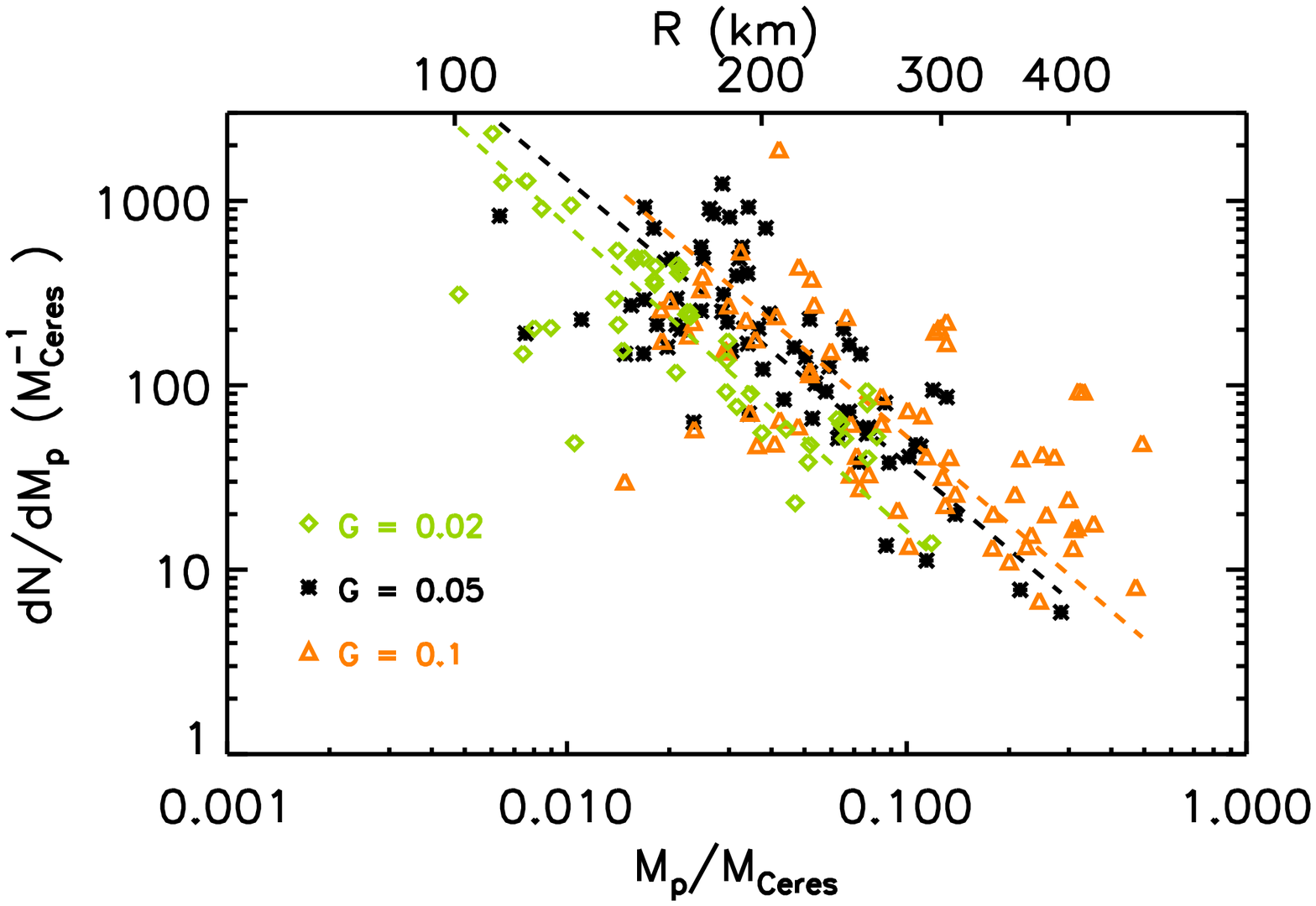}
\includegraphics[width=0.45\textwidth,angle=0]{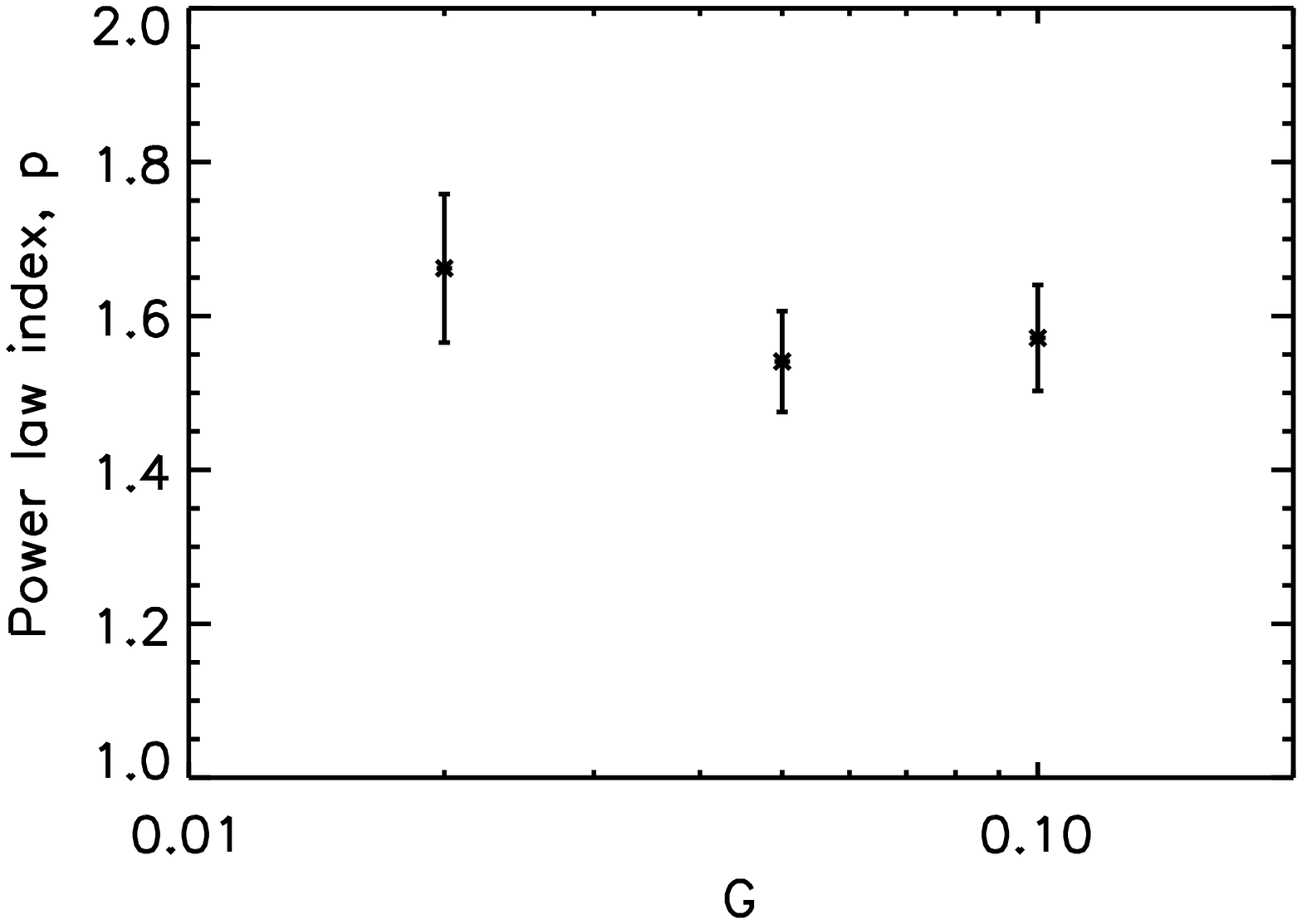}
\end{center}
\caption{Left: Differential mass distribution at $t-t_{\rm sg} = 47\Omega^{-1}$ for $\tilde{G} = 0.02$ (green diamonds), $t-t_{\rm sg} = 12.3\Omega^{-1}$ for $\tilde{G} = 0.05$ (black asterisks), and $t-t_{\rm sg} = 7.4\Omega^{-1}$ for $\tilde{G} = 0.1$ (orange triangles). Both the mass and the differential mass function are given in units of Ceres mass. On the plot, we denote the dimensionless gravity parameter with $G$ instead of $\tilde{G}$. In each case, a best fit power law is over plotted as a dashed line of the corresponding color of the data.   For each value of $\tilde{G}$, we include
the four simulations that were restarted at different times to improve statistics.  Compared to the $\tilde{G} = 0.05$ and $\tilde{G} = 0.1$ runs, a later time relative to the activation of particle self-gravity is chosen to analyze the $\tilde{G} = 0.02$ simulations because the planetesimals take a longer time to become separate, bound clumps.
Right: The best fit power law index $p$ as a function of $\tilde{G}$. The best fit slope is $p = 1.7 \pm 0.1$ for $\tilde{G} = 0.02$, $p = 1.5 \pm 0.1$  for $\tilde{G} = 0.05$, and $p = 1.6 \pm 0.1$ for $\tilde{G} = 0.1$. The mass distribution
shifts towards larger masses with increasing $\tilde{G}$.  There is no consistent trend in the value of $p$ with $\tilde{G}$.\vspace{0.3in}}
\label{dist1d_g}
\end{figure*}

As with the previous simulations, we calculate the differential mass distribution.  The result is shown in Fig.~\ref{dist1d_g}. As with the fiducial calculation, we have run an additional four simulations for each value of $\tilde{G}$
corresponding to self-gravity activated at $20\Omega^{-1}$ and $10\Omega^{-1}$ both before and after the time that it is activated in the reference simulations.  As before, this approach significantly improves
the statistical scatter in the mass distribution plot.

We examine the best fit power law slope as a function of $\tilde{G}$ in the right panel of Fig.~\ref{dist1d_g}.  The power law slope is $p = 1.7 \pm 0.1$ for $\tilde{G} = 0.02$, $p = 1.5 \pm 0.1$  for $\tilde{G} = 0.05$, and $p = 1.6 \pm 0.1$ for $\tilde{G} = 0.1$; there is some variance in $p$ with $\tilde{G}$, but these values fall within the range of $p = 1.4$--$1.8$.

\begin{figure}[!ht]
\begin{center}
\includegraphics[width=0.5\textwidth,angle=0]{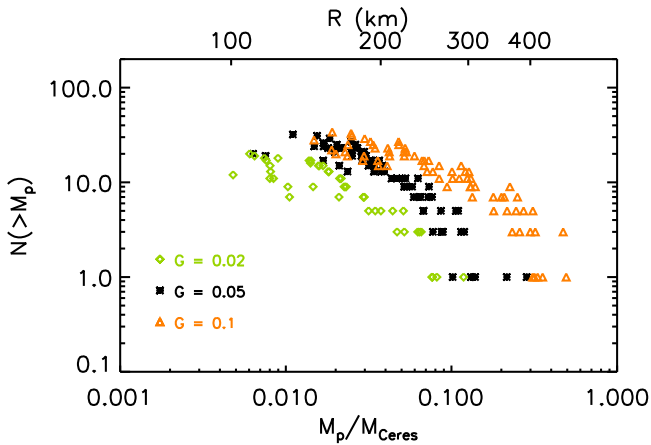}
\end{center}
\caption{Cumulative mass distribution for $\tilde{G} = 0.02$ (green diamonds),  $\tilde{G} = 0.05$ (black asterisks), and  $\tilde{G} = 0.1$ (orange triangles).   On the plot, we denote the dimensionless gravity parameter with $G$ instead of $\tilde{G}$. For each value of $\tilde{G}$, we include the four simulations that were restarted at different times. The mass distribution
shifts towards larger masses.}
\label{dist1d_g_cum}
\end{figure}

Despite the improvement introduced by combining simulation data, there is still significant scatter in the values of the differential mass function.  To remove some of this noise, we calculate the cumulative mass distribution
for each value of $\tilde{G}$ as shown in Fig.~\ref{dist1d_g_cum}.  As with the differential mass distribution, the slope of the distribution remains roughly constant with $\tilde{G}$, but the distribution shifts
towards larger mass.

\subsection{Effect of Initial Clumping}

\begin{figure*}[!ht]
\begin{center}
\includegraphics[width=0.45\textwidth,angle=0]{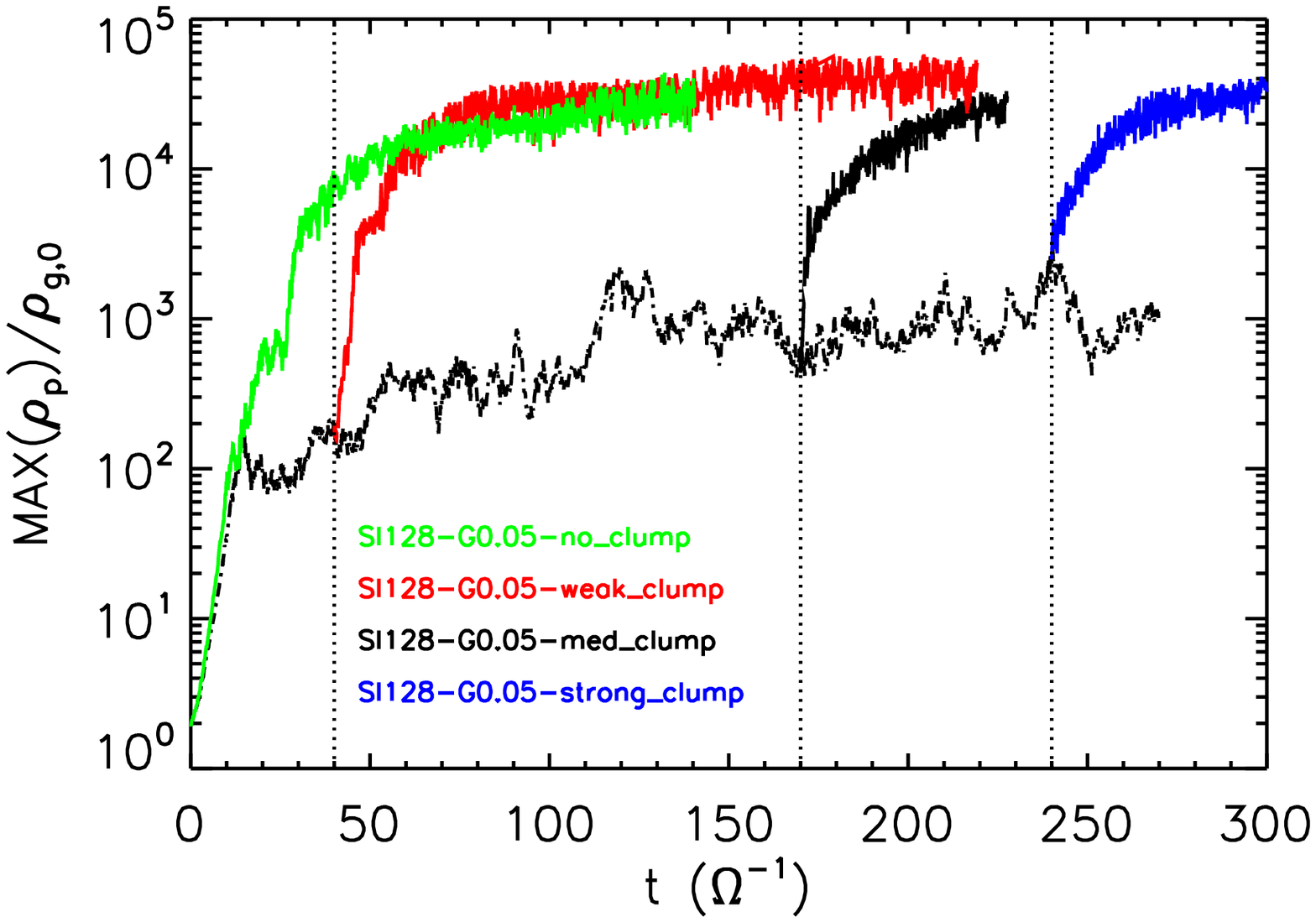}
\includegraphics[width=0.45\textwidth,angle=0]{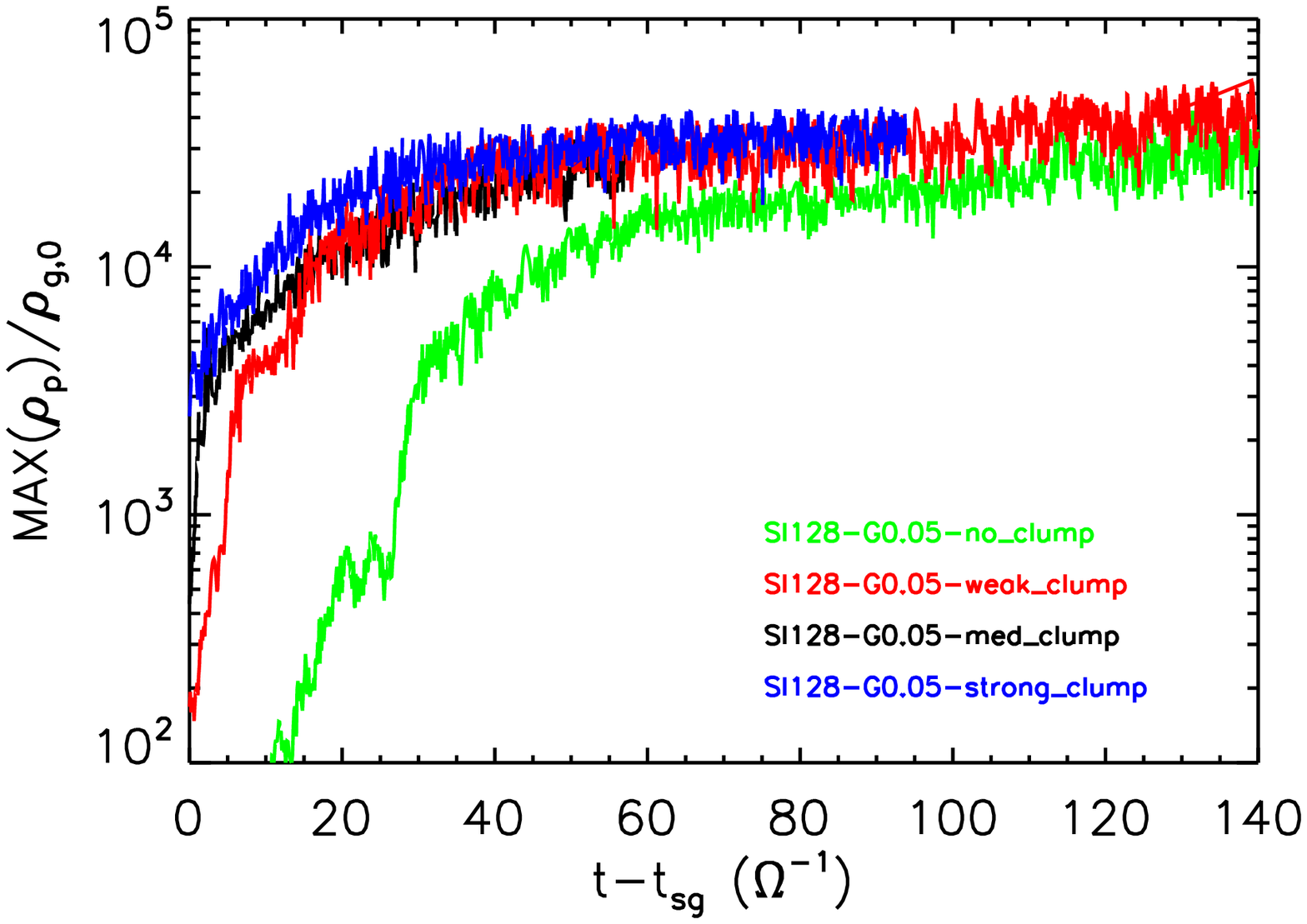}
\end{center}
\caption{The maximum of the particle mass density, normalized to the initial mid-plane gas density, as a function of time for simulations in which self-gravity was initiated at different degrees of clumping via the streaming instability. The green line
corresponds to the ``no clumping" case, SI128-G0.05-no\_clump, red is the ``weak clumping" case, SI128-G0.05-weak\_clump, black is the ``medium clumping" case, SI128-G0.05-med\_clump, and blue is the
``strong clumping" case, SI128-G0.05-strong\_clump. The left plot includes the non-self-gravitating run (dot-dashed line) from which these various simulations were restarted, whereas
the right panel displaces the temporal axis by the restart time so that the various density evolutions can be more directly compared. The growth rate of maximum particle density decreases with decreasing degree of clumping, though at late times, the maximum particle density is approximately the same between all runs.}
\label{dmax_sgstart}
\end{figure*}

In this section, we consider the effect of changing the time at which particle self-gravity is activated.  We first consider the maximum particle mass density as a function of time for this series of simulations, which
is shown in Fig.~\ref{dmax_sgstart}.  There is a clear difference in the initial growth of $\rho_{p,{\rm max}}$ depending on the initial degree of clumping from which the simulation was activated.   At late times,
the maximum particle density reaches approximately the same level for all four simulations. 

\begin{figure*}[!ht]
\begin{center}
\includegraphics[width=0.45\textwidth,angle=0]{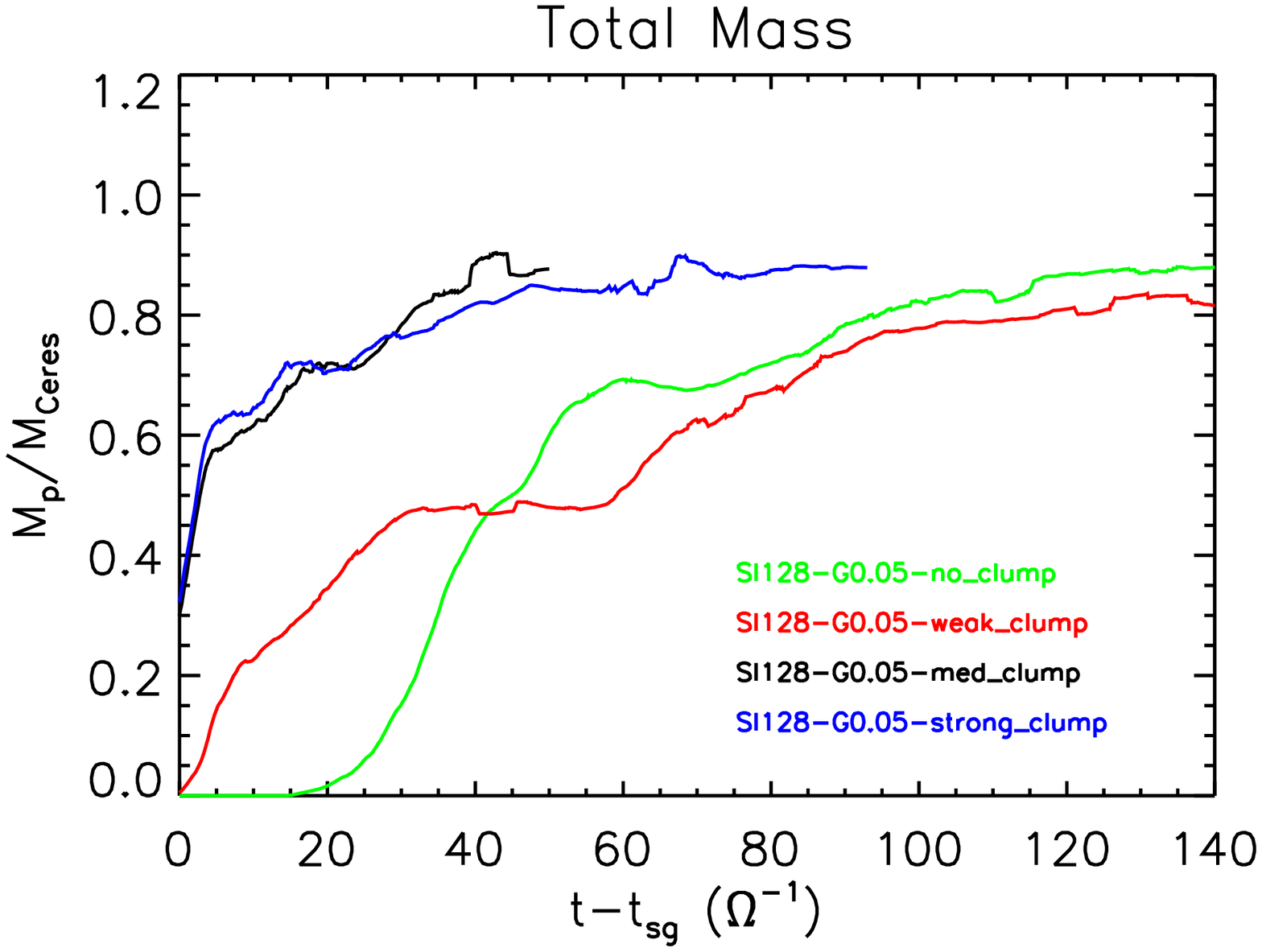}
\includegraphics[width=0.45\textwidth,angle=0]{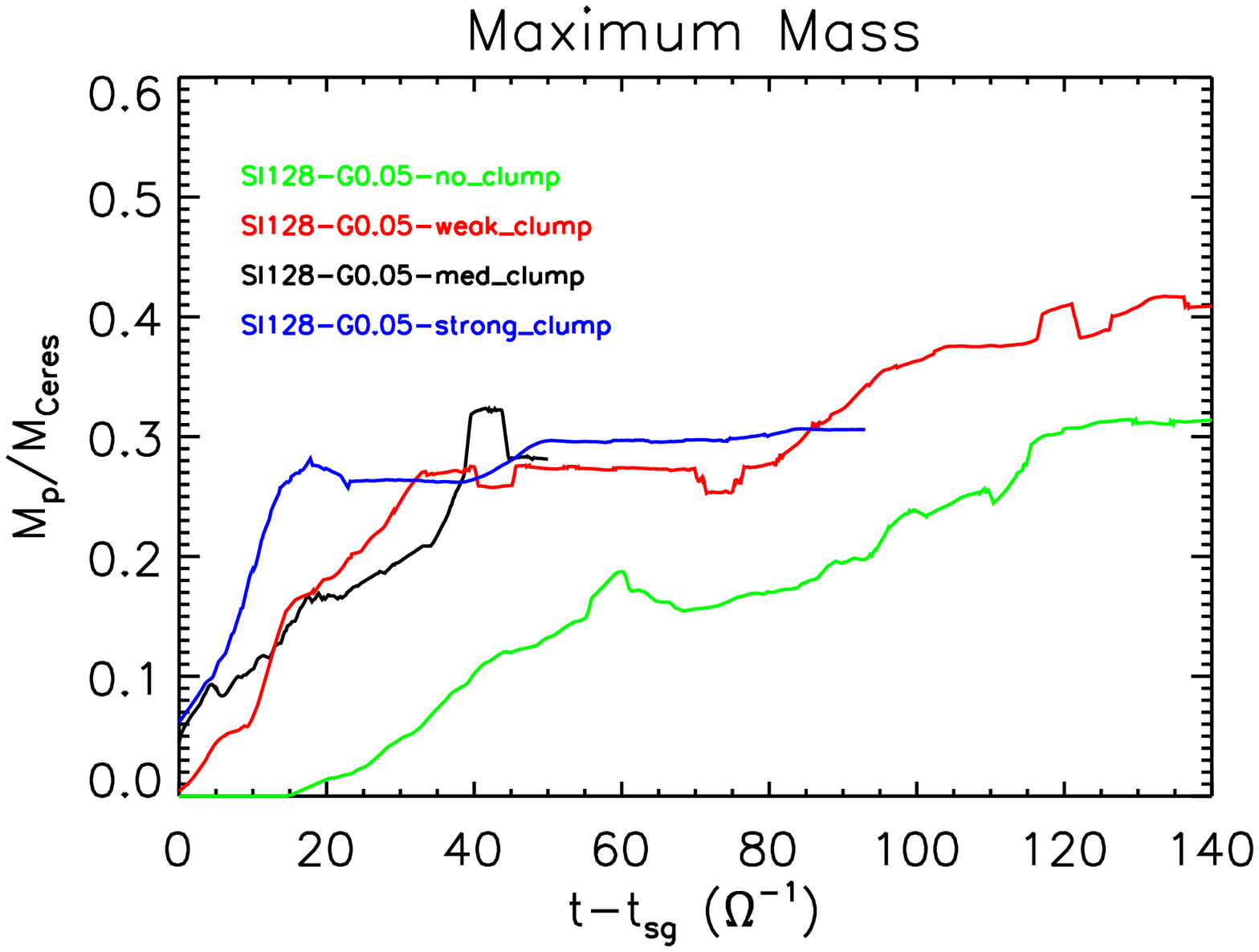}
\end{center}
\caption{The total (left) and maximum (right) planetesimal mass as a function of time for simulations in which self-gravity was initiated at different degrees of clumping via the streaming instability. The green line
corresponds to the ``no clumping" case, SI128-G0.05-no\_clump, red is the ``weak clumping" case, SI128-G0.05-weak\_clump, black is the ``medium clumping" case, SI128-G0.05-med\_clump, and blue is the
``strong clumping" case, SI128-G0.05-strong\_clump.  The growth rate of planetesimal masses decreases with decreasing degree of clumping. At late times, the masses of the four simulations become roughly equal.}
\label{sgstart_mass}
\end{figure*}

These basic results are corroborated by examination of the total and maximum planetesimal mass evolution, as shown in Fig.~\ref{sgstart_mass}. We should reiterate that
the initial stages of growth should be treated with some caution here due to the possible presence of false-positives and overlapping clumps that cause errors in our clump-finding algorithm.  However,
that the initial evolution is generally consistent with the evolution of the maximum particle density in Fig.~\ref{dmax_sgstart} is encouraging.

Given the different growth rates for the planetesimals in each of these runs, one has to be careful in choosing the correct times to analyze properties of the planetesimal distribution. Our general method has been 
the same as described above; visually examine the collapse of planetesimals and then average the mass distribution over a period corresponding to roughly when individual clumps can be identified but before there is significant merging between these clumps. 

As shown in Fig.~\ref{sgstart_mass}, there is a steep growth early on in SI128-SG-weak\_clump, followed by a relatively flat evolution, which is then followed by growth again.  This second growth period, which
happens roughly between $t-t_{\rm sg} = 60\Omega^{-1}$ and $t-t_{\rm sg} = 100\Omega^{-1}$, is due to a second phase of
planetesimal formation. From an examination of the particle surface density evolution, this second phase of planetesimal formation appears to follow the formation of another largely axisymmetric enhancement
in the particle density.  Apparently, in this particular simulation, the conditions allowed the streaming instability to act on remaining small solids in the disk after the initial formation of planetesimals.  The
second density enhancement induced by the streaming instability then went gravitationally unstable and added to the total number of planetesimals.  

In choosing our analysis time for SI128-G0.05-weak\_clump, we thus chose a time after the second period of planetesimal growth.  At the chosen times, SI128-G0.05-no\_clump produces 11 clumps, SI128-G0.05-weak\_clump produces 10, SI128-G0.05-med\_clump produces 13 and SI128-G0.05-strong\_clump produces 11.

\begin{figure*}[!ht]
\begin{center}
\includegraphics[width=0.45\textwidth,angle=0]{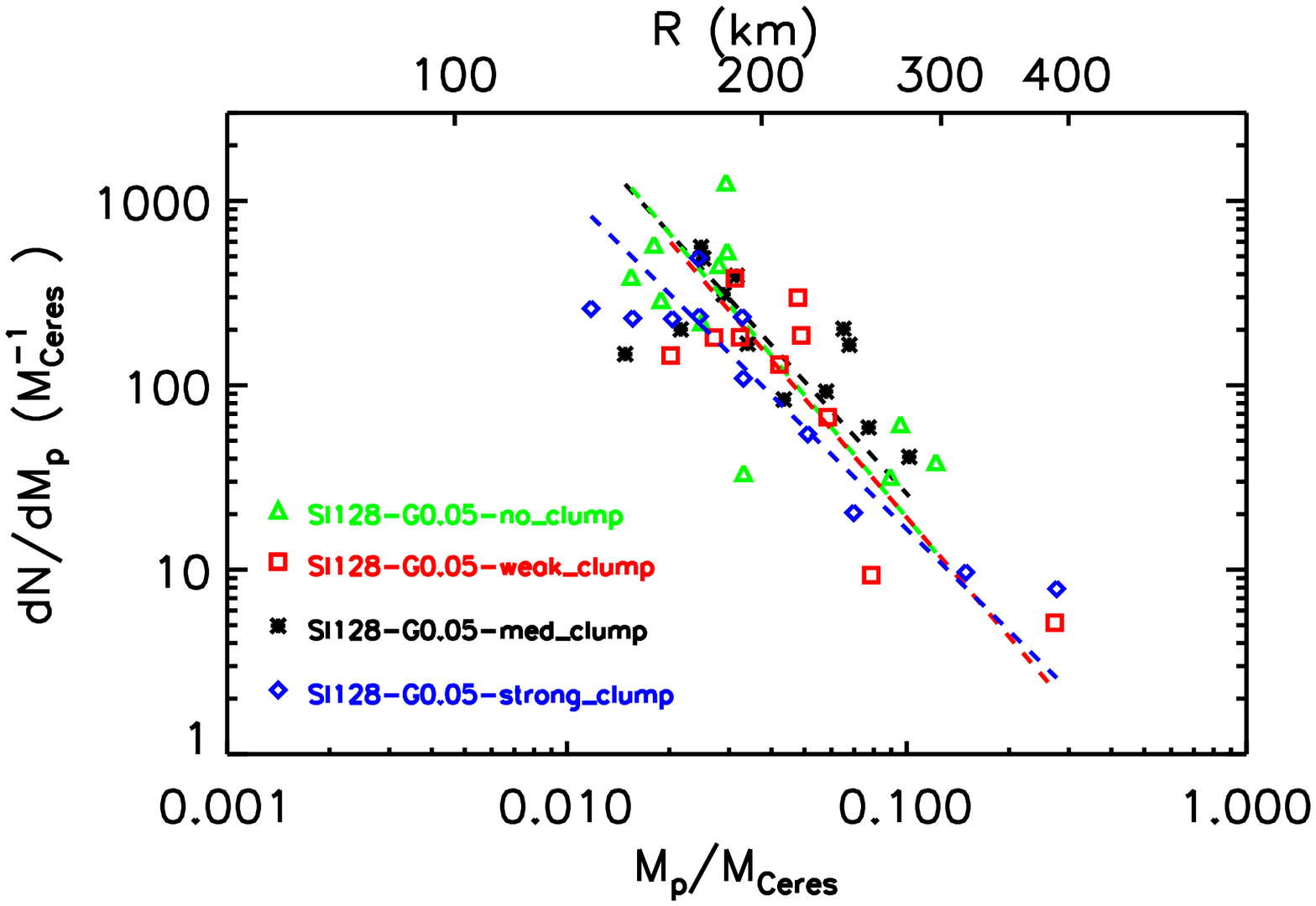}
\includegraphics[width=0.45\textwidth,angle=0]{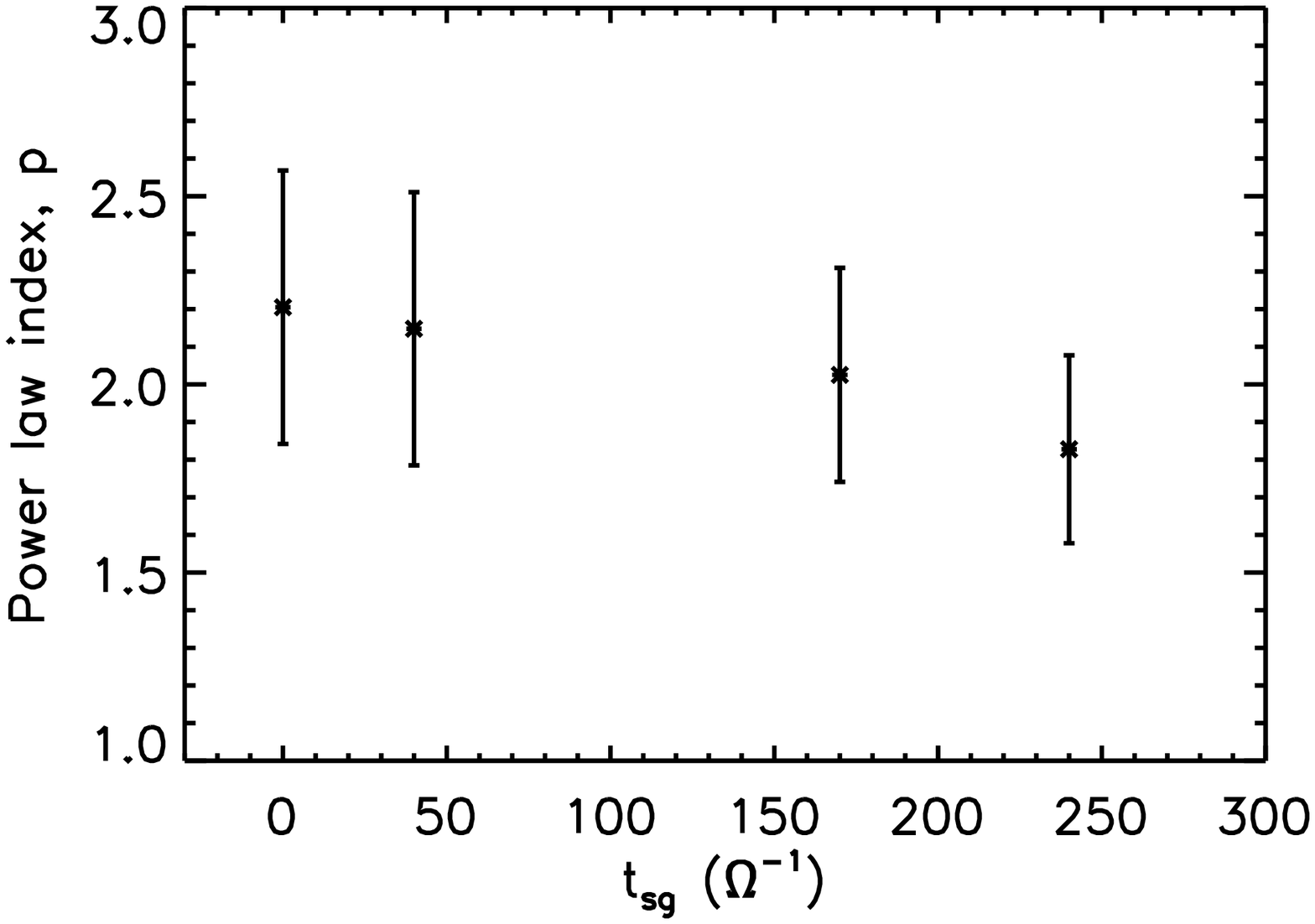}
\end{center}
\caption{Left: Differential mass distribution for simulations with self-gravity activated at different times.    The green line
corresponds to the ``no clumping" case, SI128-G0.05-no\_clump, red is the ``weak clumping" case, SI128-G0.05-weak\_clump, black is the ``medium clumping" case, SI128-G0.05-med\_clump, and blue is the
``strong clumping" case, SI128-G0.05-strong\_clump. Both the mass and the differential mass function are given in units of Ceres mass.  In each case, a best fit power law is over plotted as a dashed line of the corresponding color of the data. 
Right: The best fit power law index $p$ as a function of activation time.  The best fit values are $p = 2.2 \pm 0.4$ for SI128-G0.05-no\_clump, $p = 2.1 \pm 0.4$ for SI128-G0.05-weak\_clump, $p = 2.0 \pm 0.3$ for SI128-G0.05-med\_clump, and $p = 1.8 \pm 0.2$ for SI128-G0.05-strong\_clump. }
\label{dist1d_sgstart}
\end{figure*}

In Fig.~\ref{dist1d_sgstart}, we plot the differential mass distribution for the four different start times. There is significant scatter in the differential mass function, but the power law indices appear roughly consistent between the different simulations.  Specifically, $p = 2.2 \pm 0.4$ for SI128-G0.05-no\_clump, $p = 2.1 \pm 0.4$ for SI128-G0.05-weak\_clump, $p = 2.0 \pm 0.3$ for SI128-G0.05-med\_clump, and $p = 1.8 \pm 0.2$ for SI128-G0.05-strong\_clump.  These $p$ values are slightly larger than that found for the previously discussed simulations, which is a result
of smaller number statistics biasing the fitted slope value, as we found for the fiducial run in Section~\ref{fiducial}.

\begin{figure}[!ht]
\begin{center}
\includegraphics[width=0.45\textwidth,angle=0]{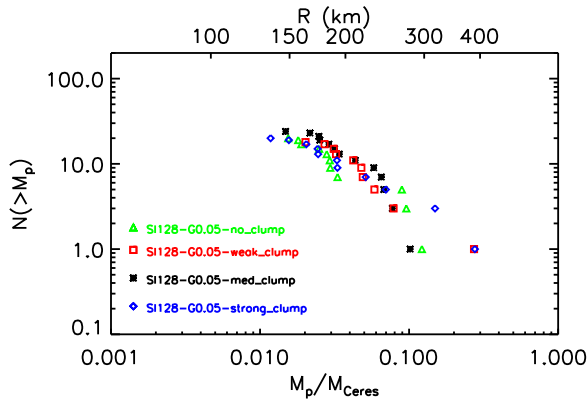}
\end{center}
\caption{Cumulative mass distribution for simulations initiated with self-gravity activated at different times    The green triangles
correspond to the ``no clumping" case, SI128-G0.05-no\_clump, red squares are the ``weak clumping" case, SI128-G0.05-weak\_clump, black asterisks are the ``medium clumping" case, SI128-G0.05-med\_clump, and blue diamonds are the
``strong clumping" case, SI128-G0.05-strong\_clump. }
\label{dist1d_sgstart_cum}
\end{figure}

As we did in Section~\ref{vary_g}, we calculate the cumulative mass distribution to reduce the noise inherent in the differential distribution.  This is shown in Fig.~\ref{dist1d_sgstart_cum}.

Roughly speaking, there does not appear to be significant differences between the mass distributions in each of these simulations.  They all occupy roughly the same space in the differential distribution plot.  Based on all of these results combined, it seems that the properties of the planetesimals do not strongly depend on the initial state from which they collapse. 

\section{Discussion}
\label{discussion}

The initial size distribution of planetesimals has a number of astrophysical implications.
Planetesimals that are too small may be subject to turbulent stirring and collisional destruction \citep{ormel13}, while planetesimals that are too large will accrete less efficiently on to giant planet core as a consequence
of less efficient gravitational focusing. While important, independent uncertainties in the modeling of these
processes make it hard to use them for quantitative constraints on planetesimal size. The best prospect
for a direct comparison between theory and observations thus remains studies of the size distribution of
small bodies in the Solar System. From our simulations we predict that $p$ generally lies in the range 1.4--1.8, which equates to a slope of the differential size distribution $q = 2.2$--$3.4$. For reasonable choices of parameters, we find that the
largest bodies in the region of the asteroid belt have masses $M_p \sim 0.1 \ M_{\rm Ceres}$, corresponding to a size of several hundred km.

The two largest populations of small bodies in the Solar System with which to compare our results are the main asteroid belt and the Kuiper belt. The asteroid belt has experienced substantial dynamical depletion (which does not alter the
sizes of surviving bodies) and collisional evolution in the time since it formed, and detailed modeling is needed to assess whether the current size distribution preserves information about the primordial population. Using such modeling,
\cite{morbidelli09} have argued that reproducing the current size distribution of asteroids {\em requires} primordial
planetesimals with diameters $D \gtrsim 100 \ {\rm km}$, and that the slope of the current size distribution above roughly
this scale is similar to the initial slope. To the extent that our work agrees with prior simulations in predicting the prompt formation of some very massive planetesimals, the \cite{morbidelli09} analysis supports the scenario of streaming-initiated
planetesimal formation. However, the power law index for the size distribution of asteroids larger than 120 km in diameter (which is relevant to our calculations) is $q \approx 4.5$ \citep{jedicke02,bottke05}. This would correspond to $p \approx 2.2$, which is significantly steeper than the power-law portion of the predicted mass function. We note, however, that if the most massive planetesimals are identified with the most massive (surviving) asteroids, then the observed slope would correspond to the region near the cut-off in the primordial mass function, and would be expected to be steeper than the power-law fit seen at lower masses.

Similar arguments apply to the Kuiper Belt. \cite{fraser10} calculated the observed luminosity function of the Kuiper belt and found a value of $q = 5.1 $ for the cold classical Kuiper belt objects and $q = 2.8$ for the excited, hot Kuiper belt population.  Later corrections to this work \citep{fraser14} that removed the model dependence associated
with uncertainties in the distance to KBOs found even steeper values of $q = 8.5$ (cold population) and $q = 5.4$ (hot population). Moreover, the largest Kuiper belt objects are bigger than those in the main asteroid belt. This is
consistent with a streaming instability model, since a
robust inference from our simulations with varying $\tilde{G}$ is that for larger gravity parameters, more massive planetesimals tend to form.   In a minimum
mass solar nebula model \citep{hayashi81}, $\tilde{G}$ slowly increases with radius (at the location of the asteroid belt, $\tilde{G} \approx 0.03$, and at 40 AU, $\tilde{G} \approx 0.05$), suggesting that
larger planetesimals should be formed at further distances from the Sun. However, as with the asteroid belt, the
observed slopes are not what we predict for the primordial population.

The above comparisons are suggestive but should be regarded as preliminary. In addition to the uncertainties in deciphering what the current size distributions in the asteroid and Kuiper belts tell us about the primordial population,
there are several sources of possible uncertainty in the prediction of the size distribution itself.  First, at a purely numerical level, the collapse of planetesimals in our simulations is halted at the grid scale, which is unphysically large. The cross-section for subsequent accretion or mergers is therefore boosted, which could impact the measured size distribution. We note, however, that the recent simulations of \cite{johansen15} --- in which collapsed planetesimals were replaced with sink particles --- yielded roughly consistent values of $p$ and $q$ as compared to our $512^3$ simulation.  Thus, it would seem
that there are no systematic errors in the determination of $p$ with high resolution simulations. 

 The largest scales in the box may also play an important role in properties of collapsed planetesimals.  Indeed, it has been recently shown through non-self-gravitating simulations that the domain size influences the temporal and spatial properties of particle clumping during the non-linear state of the streaming instability  \citep{yang14}.  It is entirely conceivable that such effects will influence the outcome of
 the gravitational collapse phase, and we are currently pursuing such investigations (Li et al., in prep).

Numerical effects aside, however, it is possible that the size distribution of planetesimals formed from gravitational collapse is not universal, and may be different in the physical environment relevant to the asteroid or Kuiper belts. The streaming instability can produce strong clumping of solids across a wide region in the parameter space of metallicity, radial
gas pressure gradient and particle size \citep{bai10b,carrera15}. These parameters are expected to vary substantially with disk radius (for example, the dimensionless stopping time for particles in the asteroid belt region is likely to be much smaller than for the Kuiper Belt region), and this may result in different size distributions after gravitational collapse. 

It is also conceivable that the size distribution is a function of how unstable the disk is to the streaming instability and gravitational collapse. We have simulated a system in which a significant fraction of the total mass of solids forms planetesimals on a time scale of just $\sim 10^2 \ \Omega^{-1}$. Meteoritic evidence from the asteroid belt, conversely, suggests a broad spread in the formation times of primitive material \citep{villeneuve09}. This can be interpreted as implying that planetesimal precursors persist in the disk over time scales more akin to $10^7 \ \Omega^{-1}$.  It is not known whether a marginally unstable system would form the same planetesimal mass function as the one that we have simulated.

Furthermore, it is worth noting that the size distribution at the end of the gravitational collapse can be subsequently modified by the longer term accretion of planetesimals and smaller solids.  The effects of this
accretion depend on the evolving size distribution of solids and the evolution of the gas, including its turbulent state \cite[e.g.,][]{johansen15}.  However, the size distribution at the end of the collapse
phase, as studied here, is important as input for these longer term accretion studies.


\section{Conclusions}
\label{conclusions}

We have developed and tested a module for the mutual gravitational interaction of particles in the {\sc Athena} code and
applied it to preliminary studies of the streaming instability under the influence of particle gravity.   As this is the
first study of planetesimal formation with the {\sc Athena} code, we have carried out a basic parameter sweep
in order to provide a baseline of calculations and results from which to spawn further investigations and 
with which to compare the results already present in the literature.

In this paper, we have varied the numerical resolution, the relative strength of gravity and tidal effects, and
the degree of clumping induced by the streaming instability before the activation of particle self-gravity.  While we 
will follow up this study in future papers to explore more parameters, we can draw some preliminary conclusions
from this work.

\begin{enumerate}

\item The streaming instability leads to enhanced particle clumping, after which the mutual gravitational attraction between solid particles leads to the formation of a number of bound planetesimals.

\item For the choice of metallicity and stopping time used here, the masses of these planetesimals range from $\sim 0.001 M_{\rm Ceres}$ to $\sim 0.1 M_{\rm Ceres}$. The typical radii of these planetesimals are 50 km to a few hundred km.  

\item Where a direct comparison is possible, we find excellent agreement between planetesimal properties in our {\sc Athena} simulations and those carried out with the {\sc Pencil} code.

\item As resolution is increased, more planetesimals are produced at lower masses and smaller radii, while the high end mass of the distribution remains approximately the same.  There is no significant trend of the power law index
$p$ with resolution.  The values of $p$ fall within the range $p \approx 1.4$--$1.7$. 

\item The power law slope of the size distribution for the highest resolution simulation is $q = 2.8$, which is significantly shallower than that measured for both the main asteroid belt ($q \approx 4.5$) and the classical Kuiper belt (cold population;
$q \approx 8.5$ hot population; $q \approx 5.4$).

\item Varying the relative strength of gravity through the parameter $\tilde{G}$ changes the mass range of planetesimals produced.  The value of $p$ does not appear to have any consistent trend with $\tilde{G}$.  

\item The properties of planetesimals appear largely independent of the initial degree of clumping before gravity takes over, justifying the method employed here and in various papers by Johansen and co-authors of activating particle self-gravity
after the streaming instability has already produced clumps.

\end{enumerate}

As this is an initial step into planetesimal studies with {\sc Athena}, we have only explored a few parameters.  A more complete study of planetesimal properties
will require varying other parameters, both physical (e.g., metallicity, particle size, radial pressure gradient) and numerical (e.g., boundary conditions and domain size).  We will
address these issues in future publications. 

A more serious uncertainty lies in the relatively small mass and size range produced by the moderate resolution
$128^3$ simulations, which generally spans only a decade in mass. While we were able to alleviate the resulting issue
of small number statistics associated with this limited mass range by combining datasets, the small mass
end of the distribution is ultimately determined by the finite grid scale.  Going to higher resolution and a larger number of particles
is the obvious solution to this problem, and indeed the highest resolution $512^3$ simulation spanned
nearly two mass decades.  However, such a simulation is very computationally expensive and this expense makes a large exploration of parameter space infeasible. 

In a similar vein, at any given resolution, we do not resolve compact planetesimals; i.e., the planetesimals that do form
are large compared to the size of planetesimals that would form if we resolved particle self-gravity below the grid scale. These unphysical sizes can
cause enhanced accretion of smaller solids onto formed planetesimals and make the power law distributions shallower.  The 
future implementation of sink particles or sub-grid-cell gravity will help to alleviate these issues.

Despite these uncertainties, our first principles calculations provide additional evidence that the streaming instability can form planetesimals with properties consistent with Solar System constraints. The largest remaining source of uncertainty may be in the initial metallicity and particle size distribution (i.e., {\em prior} to the onset of the streaming instabilty). If the appropriate initial conditions can be pinned down via astronomical observations, more comprehensive numerical simulations should be able to provide a detailed understanding of the formation and evolution of the planetesimal population.


\acknowledgements
We thank Daniel R. Wik for his handy knowledge of statistics and Katherine Kretke for many discussions on observational constraints of Solar System planetesimal populations.  
We also thank Hal Levison, Anders Johansen, Xue-Ning Bai, and Chao-Chin Yang for useful discussions regarding this work and the referee, whose suggestions greatly improved
the quality of this work. We acknowledge support from NASA through grants NNX13AI58G  and NNX16AB42G (P.J.A), from the NSF through grant 
AST 1313021 (P.J.A.), and from grant HST-AR-12814 (P.J.A.) awarded by the Space Telescope Science Institute, which is operated by the  Association of Universities for Research
in Astronomy, Inc., for NASA, under contact NAS 5-26555. J.B.S.'s support was provided in part under contract with
the California Institute of Technology (Caltech) and the Jet Propulsion Laboratory (JPL) funded by NASA through the Sagan Fellowship Program executed by the
NASA Exoplanet Science Institute.  The computations were performed on Stampede and Maverick at the Texas Advanced Computing Center through XSEDE grant TG-AST120062.


\begin{thebibliography}{43}
\expandafter\ifx\csname natexlab\endcsname\relax\def\natexlab#1{#1}\fi

\bibitem[{Andrews {et~al.}(2012)Andrews, Wilner, Hughes, Qi, Rosenfeld,
  {\"O}berg, Birnstiel, Espaillat, Cieza, Williams, Lin, \& Ho}]{andrews12}
Andrews, S.~M., Wilner, D.~J., Hughes, A.~M., {et~al.} 2012, The Astrophysical
  Journal, 744, 162

\bibitem[{Bai \& Stone(2010{\natexlab{a}})}]{bai10c}
Bai, X.-N., \& Stone, J.~M. 2010{\natexlab{a}}, The Astrophysical Journal, 722,
  1437

\bibitem[{Bai \& Stone(2010{\natexlab{b}})}]{bai10a}
---. 2010{\natexlab{b}}, The Astrophysical Journal Supplement, 190, 297

\bibitem[{Bai \& Stone(2010{\natexlab{c}})}]{bai10b}
---. 2010{\natexlab{c}}, The Astrophysical Journal Letters, 722, L220

\bibitem[{Birnstiel {et~al.}(2010)Birnstiel, Dullemond, \&
  Brauer}]{birnstiel10}
Birnstiel, T., Dullemond, C.~P., \& Brauer, F. 2010, Astronomy and Astrophysics

\bibitem[Birnstiel et 
al.(2012)]{birnstiel12} Birnstiel, T., Andrews, S.~M., \& Ercolano, B.\ 2012, \aap, 544, A79 

\bibitem[Bitsch et 
al.(2014)]{bitsch14} Bitsch, B., Morbidelli, A., Lega, E., Kretke, K., \& Crida, A.\ 2014, \aap, 570, A75 

\bibitem[{Blum \& Wurm(2008)}]{blum08}
Blum, J., \& Wurm, G. 2008, Annual Review of Astronomy and Astrophysics, 46, 21

\bibitem[{Bottke {et~al.}(2005)Bottke, Durda, Nesvorn{\'{y}}, Jedicke,
  Morbidelli, Vokrouhlick{\'{y}}, \& Levison}]{bottke05}
Bottke, W.~F., Durda, D.~D., Nesvorn{\'{y}}, D., {et~al.} 2005, Icarus, 175,
  111

\bibitem[{Brauer {et~al.}(2008)Brauer, Dullemond, \& Henning}]{brauer08}
Brauer, F., Dullemond, C.~P., \& Henning, T. 2008, Astronomy and Astrophysics,
  480, 859

\bibitem[{Carrera {et~al.}(2015)Carrera, Johansen, \& Davies}]{carrera15}
Carrera, D., Johansen, A., \& Davies, M.~B. 2015, arXiv.org

\bibitem[{Clauset {et~al.}(2009)Clauset, Shalizi, \& Newman}]{clauset09}
Clauset, A., Shalizi, C.~R., \& Newman, M. E.~J. 2009, SIAM Review

\bibitem[{Colella(1990)}]{colella90}
Colella, P. 1990, JCP, 87, 171

\bibitem[{Colella \& Woodward(1984)}]{colella84}
Colella, P., \& Woodward, P.~R. 1984, JCP, 54, 174

\bibitem[Dr{\c a}{\.z}kowska et 
al.(2013)]{drazkowska13} Dr{\c a}{\.z}kowska, J., Windmark, F., \& Dullemond, C.~P.\ 2013, \aap, 556, A37 

\bibitem[Dr{\c a}{\.z}kowska \& Dullemond(2014)]{drazkowska14} Dr{\c a}{\.z}kowska, J., \& Dullemond, C.~P.\ 2014, \aap, 572, A78 

\bibitem[{Dzyurkevich {et~al.}(2010)Dzyurkevich, Flock, Turner, Klahr, \&
  Henning}]{dzyurkevich10}
Dzyurkevich, N., Flock, M., Turner, N.~J., Klahr, H., \& Henning, T. 2010,
  A\&A, 515, 70

\bibitem[{Fraser {et~al.}(2014)Fraser, Brown, Morbidelli, Parker, \&
  Batygin}]{fraser14}
Fraser, W.~C., Brown, M.~E., Morbidelli, A., Parker, A., \& Batygin, K. 2014,
  The Astrophysical Journal, 782, 100

\bibitem[{Fraser {et~al.}(2010)Fraser, Brown, \& Schwamb}]{fraser10}
Fraser, W.~C., Brown, M.~E., \& Schwamb, M.~E. 2010, Icarus, 210, 944

\bibitem[{Gardiner \& Stone(2005)}]{gardiner05a}
Gardiner, T.~A., \& Stone, J.~M. 2005, JCP, 205, 509

\bibitem[{Gardiner \& Stone(2008)}]{gardiner08}
---. 2008, JCP, 227, 4123

\bibitem[{Hawley {et~al.}(1995)Hawley, Gammie, \& Balbus}]{hawley95a}
Hawley, J.~F., Gammie, C.~F., \& Balbus, S.~A. 1995, ApJ, 440, 742

\bibitem[{Hayashi(1981)}]{hayashi81}
Hayashi, C. 1981, Progress of Theoretical Physics Supplement, 70, 35

\bibitem[{Jedicke {et~al.}(2002)Jedicke, Larsen, \& Spahr}]{jedicke02}
Jedicke, R., Larsen, J., \& Spahr, T. 2002, Asteroids III, 71

\bibitem[{Johansen {et~al.}(2011)Johansen, Klahr, \& Henning}]{johansen11a}
Johansen, A., Klahr, H., \& Henning, T. 2011, Astronomy and Astrophysics, 529,
  A62

\bibitem[{Johansen {et~al.}(2015)Johansen, Mac~Low, Lacerda, \&
  Bizzarro}]{johansen15}
Johansen, A., Mac~Low, M.-M., Lacerda, P., \& Bizzarro, M. 2015, Science
  Advances, 1, 1500109

\bibitem[{Johansen {et~al.}(2007)Johansen, Oishi, Mac~Low, Klahr, Henning, \&
  Youdin}]{johansen07a}
Johansen, A., Oishi, J.~S., Mac~Low, M.-M., {et~al.} 2007, Nature, 448, 1022

\bibitem[{Johansen \& Youdin(2007)}]{johansen07b}
Johansen, A., \& Youdin, A. 2007, The Astrophysical Journal, 662, 627

\bibitem[{Johansen {et~al.}(2009{\natexlab{a}})Johansen, Youdin, \&
  Klahr}]{johansen09a}
Johansen, A., Youdin, A., \& Klahr, H. 2009{\natexlab{a}}, The Astrophysical
  Journal, 697, 1269

\bibitem[{Johansen {et~al.}(2009{\natexlab{b}})Johansen, Youdin, \&
  Mac~Low}]{johansen09c}
Johansen, A., Youdin, A., \& Mac~Low, M.-M. 2009{\natexlab{b}}, The
  Astrophysical Journal Letters, 704, L75

\bibitem[{Johansen {et~al.}(2012)Johansen, Youdin, \& Lithwick}]{johansen12}
Johansen, A., Youdin, A.~N., \& Lithwick, Y. 2012, Astronomy and Astrophysics,
  537, A125

\bibitem[{Johnson {et~al.}(2008)Johnson, Guan, \& Gammie}]{johnson08}
Johnson, B.~M., Guan, X., \& Gammie, C.~F. 2008, ApJS, 179, 553

\bibitem[Kataoka et 
al.(2013)]{kataoka13} Kataoka, A., Tanaka, H., Okuzumi, S., \& Wada, K.\ 2013, \aap, 557, L4 

\bibitem[{Koyama \& Ostriker(2009)}]{koyama09}
Koyama, H., \& Ostriker, E.~C. 2009, The Astrophysical Journal, 693, 1316

\bibitem[{Kretke \& Lin(2007)}]{kretke07}
Kretke, K.~A., \& Lin, D. N.~C. 2007, ApJ, 664, L55

\bibitem[Krijt et 
al.(2015)]{krijt15} Krijt, S., Ormel, C.~W., Dominik, C., \& Tielens, A.~G.~G.~M.\ 2015, \aap, 574, A83 

\bibitem[{Masset(2000)}]{masset00}
Masset, F. 2000, A\&AS, 141, 165

\bibitem[{Morbidelli {et~al.}(2009)Morbidelli, Bottke, Nesvorn{\'{y}}, \&
  Levison}]{morbidelli09}
Morbidelli, A., Bottke, W.~F., Nesvorn{\'{y}}, D., \& Levison, H.~F. 2009,
  Icarus, 204, 558
  
  \bibitem[Musiolik et al.(2016)]{musiolik16} Musiolik, G., Teiser, 
J., Jankowski, T., \& Wurm, G.\ 2016, \apj, 818, 16 

\bibitem[Okuzumi et al.(2012)]{okuzumi12} Okuzumi, S., Tanaka, 
H., Kobayashi, H., \& Wada, K.\ 2012, \apj, 752, 106 

\bibitem[{Ormel \& Okuzumi(2013)}]{ormel13}
Ormel, C.~W., \& Okuzumi, S. 2013, The Astrophysical Journal, 771, 44

\bibitem[Sekora 
\& Colella(2009)]{sekora09} Sekora, M., \& Colella, P.\ 2009, arXiv:0903.4200 

\bibitem[{Simon {et~al.}(2011)Simon, Hawley, \& Beckwith}]{simon11a}
Simon, J.~B., Hawley, J.~F., \& Beckwith, K. 2011, ApJ, 730, 94

\bibitem[Simon 
\& Armitage(2014)]{simon14} Simon, J.~B., \& Armitage, P.~J.\ 2014, \apj, 784, 15 

\bibitem[{Stone \& Gardiner(2010)}]{stone10}
Stone, J.~M., \& Gardiner, T.~A. 2010, ApJS, 189, 142

\bibitem[{Stone {et~al.}(2008)Stone, Gardiner, Teuben, Hawley, \&
  Simon}]{stone08}
Stone, J.~M., Gardiner, T.~A., Teuben, P., Hawley, J.~F., \& Simon, J.~B. 2008,
  The Astrophysical Journal Supplement, 178, 137

\bibitem[{Toro(1999)}]{toro99}
Toro, E.~F. 1999, {Riemann solvers and numerical models for fluid dynamics}

\bibitem[Villeneuve et al.(2009)]{villeneuve09} Villeneuve, J., 
Chaussidon, M., \& Libourel, G.\ 2009, Science, 325, 985 

\bibitem[Wada et 
al.(2013)]{wada13} Wada, K., Tanaka, H., Okuzumi, S., et al.\ 2013, \aap, 559, A62 

\bibitem[{Weidenschilling(1977)}]{weidenschilling77b}
Weidenschilling, S.~J. 1977, MNRAS, 180, 57

\bibitem[Yang 
\& Menou(2010)]{yang10} Yang, C.-C., \& Menou, K.\ 2010, \mnras, 402, 2436 

\bibitem[Yang \& Johansen(2014)]{yang14} Yang, C.-C., \& Johansen, A.\ 2014, \apj, 792, 86

\bibitem[{Youdin \& Johansen(2007)}]{youdin07a}
Youdin, A., \& Johansen, A. 2007, The Astrophysical Journal, 662, 613

\bibitem[{Youdin \& Goodman(2005)}]{youdin05}
Youdin, A.~N., \& Goodman, J. 2005, The Astrophysical Journal, 620, 459

\bibitem[{Zsom {et~al.}(2010)Zsom, Ormel, G{\"u}ttler, Blum, \&
  Dullemond}]{zsom10}
Zsom, A., Ormel, C.~W., G{\"u}ttler, C., Blum, J., \& Dullemond, C.~P. 2010,
  Astronomy and Astrophysics, 513, A57

\end{thebibliography}
\end{document}